\documentclass[12pt, peerreview , onecolumn]{IEEEtran}
\pdfoutput=1

\usepackage{tikz}
\usepackage[sumlimits,intlimits]{amsmath}
\usepackage{amsmath,amsfonts,amssymb}%
\usepackage{bm}%
\usepackage{graphicx,graphics}%

\usepackage{cases}%
\usepackage[noadjust]{cite}%
\usepackage{color}%
\usepackage{cite,url}
\usepackage{verbatim}
\usepackage{algorithm}
\usepackage{balance}
\usepackage{multirow}
\usepackage{multicol}
\usepackage{stfloats}
\usepackage{subfigure}
\usepackage{xspace}
\usepackage{enumerate}
\usepackage{rotating}

\ifCLASSOPTIONpeerreview

\fi

\newcommand{\lh}{\lambda_h}
\newcommand{\lgg}{\lambda_g}
\newcommand{\sigr}{\sigma^2_{n^{[r]}}}
\newcommand{\sigra}{\sigma^2_{n_a^{[r]}}}
\newcommand{\sigrp}{\sigma^2_{n_c^{[r]}}}
\newcommand{\sigd}{\sigma^2_{n^{[d]}}}
\newcommand{\sigda}{\sigma^2_{n_a^{[d]}}}
\newcommand{\sigdp}{\sigma^2_{n_c^{[d]}}}
\newcommand{\siga}{\sigma^2_{n_a}}
\newcommand{\sigp}{\sigma^2_{n_c}}
\newcommand{\pout}{p_\text{out}}

\begin{document}

\ifCLASSOPTIONpeerreview
    \title{\vspace*{-0pt}{ Relaying Protocols for Wireless Energy}
\\ \vspace*{-20pt}
Harvesting and Information Processing}
\else
\title{ Relaying Protocols for Wireless Energy Harvesting and Information Processing}
\fi

\author{\IEEEauthorblockN{Ali A.~Nasir, \textit{Student Member, IEEE}, Xiangyun Zhou, \textit{Member, IEEE}}, \\ Salman Durrani, \textit{Senior Member, IEEE}, and Rodney A. Kennedy, \textit{Fellow, IEEE}}

\maketitle

{\let\thefootnote\relax\footnotetext{ Ali A. Nasir, Xiangyun Zhou, Salman Durrani, and Rodney Kennedy are with Research School of Engineering, The Australian National University, Canberra, Australia.  Emails: ali.nasir@anu.edu.au, xiangyun.zhou@anu.edu.au, salman.durrani@anu.edu.au, and rodney.kennedy@anu.edu.au. This research was supported under Australian Research Council's Discovery Projects funding scheme (project number DP110102548). \vspace{-00pt}}}

\vspace{-12pt}

\begin{abstract}
An emerging solution for prolonging the lifetime of energy constrained relay nodes in wireless networks is to avail the ambient radio-frequency (RF) signal and to simultaneously harvest energy and process information. In this paper, an amplify-and-forward (AF) relaying network is considered, where an energy constrained relay node harvests energy from the received RF signal and uses that harvested energy to forward the source information to the destination. Based on the time switching and power splitting receiver architectures, two relaying protocols, namely, i) time switching-based relaying (TSR) protocol and ii) power splitting-based relaying (PSR) protocol are proposed to enable energy harvesting and information processing at the relay.  In order to determine the throughput, analytical expressions for the outage probability and the ergodic capacity are derived for delay-limited and delay-tolerant transmission modes, respectively. The numerical analysis provides practical insights into the effect of various system parameters, such as energy harvesting time, power splitting ratio, source transmission rate, source to relay distance, noise power, and energy harvesting efficiency, on the performance of wireless energy harvesting and information processing using AF relay nodes. In particular, the TSR protocol outperforms the PSR protocol in terms of throughput at relatively low signal-to-noise-ratios and high transmission rate.

\end{abstract}


\ifCLASSOPTIONpeerreview
    \newpage
\fi

\section{Introduction}

Prolonging the lifetime of a wireless network through energy harvesting has received significant attention very recently \cite{Xu-12-A,Ho-12-A,Luo-12-A,Varshney-08-P,Grover-10-P,Zhou-12-A,Fouladgar-12-A,Chalise-12-P,Huang-12-A,Popovski-12-A,Liu-12-A,Zhang-12-A}.
Though, replacing or recharging batteries can avoid energy harvesting, it incurs a high cost and can be inconvenient or hazardous (e.g., in a toxic environments), or highly undesirable (e.g., for sensors embedded in building structures or inside the human body) \cite{Zhang-12-A}. In such scenarios, a safe and convenient option may be to harvest the energy from the environment. Apart from the conventional energy harvesting methods, such as solar, wind, vibration, thermoelectric effects or other physical phenomena \cite{Raghunathan-06-A,Paradiso-05-A,Xu-12-A,Ho-12-A,Luo-12-A,Huang-12-A,Medepally-10-A}, a new emerging solution is to avail ambient radio-frequency (RF) signals \cite{Varshney-08-P}. The advantage of this solution lies in the fact that RF signals can carry energy and information at the same time. Thus, energy constrained nodes can scavenge energy and process the information simultaneously \cite{Zhang-12-A,Varshney-08-P,Grover-10-P,Zhou-12-A,Fouladgar-12-A,Popovski-12-A}.

For wireless energy harvesting using RF signals, the recent state-of-the-art advances in point-to-point systems can be classified into two main approaches. The first approach considers an ideal receiver design that is able to simultaneously observe and extract power from the same received signal \cite{Varshney-08-P,Grover-10-P,Popovski-12-A}. However, as discussed in \cite{Zhou-12-A}, this assumption does not hold in practice, as practical circuits for harvesting energy from RF signals are not yet able to decode the carried information directly. The second approach considers a practically realizable receiver design with separate information decoding and energy harvesting receiver for information and power transfer and is now widely adopted in the literature \cite{Zhou-12-A,Xiang-12-A,Zhang-12-A,Liu-12-A,Lee-12-A}. For the first class of receivers, the idea of transmitting information and energy simultaneously was first proposed in \cite{Varshney-08-P}, where the authors used a capacity-energy function to study the fundamental performance tradeoff for simultaneous information and power transfer. The work in \cite{Varshney-08-P} was extended to frequency-selective channels with additive white Gaussian noise (AWGN) in \cite{Grover-10-P}. A two-way communication system for energy harvesting and information transmission was investigated in \cite{Popovski-12-A}. For the second class of receivers, the performance limits of a three node multiple-input-multiple-output (MIMO) broadcasting system, with separate energy harvesting and information decoding receiver, was studied in \cite{Zhang-12-A}. The work in \cite{Zhang-12-A} was extended in \cite{Xiang-12-A} by considering imperfect channel state information (CSI) at the transmitter. Subject to co-channel interference, optimal designs to achieve different outage-energy and rate-energy tradeoffs in delay-limited and delay-tolerant transmission modes were formulated in \cite{Liu-12-A}. The application of wireless energy harvesting to a cognitive radio network was considered in \cite{Lee-12-A}, where the throughput of the secondary network was maximized under an outage constraint for primary and secondary networks.


\subsection{Motivation and Contribution}

The majority of the recent research in wireless energy harvesting and information processing has considered point-to-point communication systems \cite{Varshney-08-P,Grover-10-P,Liu-12-A,Zhou-12-A,Popovski-12-A,Zhang-12-A,Xiang-12-A,Lee-12-A}. In wireless cooperative or sensor networks, the relay or sensor nodes may have limited battery reserves and may need to rely on some external charging mechanism in order to remain active in the network \cite{Medepally-10-A,Venkata-12-A}. Therefore, energy harvesting in such networks is particularly important as it can enable information relaying.

In this paper, we are concerned with the problem of wireless energy harvesting and information processing in an amplify-and-forward (AF) wireless cooperative or sensor network. We consider the scenario that an energy constrained relay node harvests energy from the RF signal broadcasted by a source node and uses that harvested energy to forward the source signal to a destination node. We adopt time switching (TS) and power splitting (PS) receiver architectures, as proposed in \cite{Zhou-12-A}. Based on the receiver architectures in \cite{Zhou-12-A} and the well-known AF relaying protocol \cite{Laneman-04-A}, we propose two relaying protocols i) TS-based relaying (TSR) protocol and ii) PS-based relaying (PSR) protocol for separate information processing and energy harvesting at the energy constrained relay node. In TSR protocol, the relay spends some time for energy harvesting and the remaining time for information processing. In PSR protocol, the relay uses a portion of the received power for energy harvesting and the remaining power for information processing. Our figure of merit is the throughput, which is defined as the number of bits that are successfully decoded per unit time per unit bandwidth at the destination node. We formulate and study the throughput for both the TSR and the PSR protocols with delay-limited and delay-tolerant transmission modes, where outage probability and ergodic capacity are derived to evaluate the throughput in delay-limited and delay-tolerant transmission modes, respectively. Finally, we also derive the achievable throughput of an ideal relay receiver, that is based on the ideal receiver in \cite{Varshney-08-P,Grover-10-P,Popovski-12-A}, and processes the information and extracts power from the same received signal \cite{Zhou-12-A}. The main contributions of this paper are summarized as follows:

\begin{itemize}
\item We propose the TSR and the PSR protocols to enable wireless energy harvesting and information processing at the energy constrained relay in wireless AF relaying networks, based on the TS and PS receiver architectures.
\item For the TSR and the PSR protocols, we derive analytical expressions for the achievable throughput at the destination by i) evaluating the outage probability for delay-limited transmission mode and ii) evaluating the ergodic capacity for delay-tolerant transmission mode. The derived expressions provide practical design insights into the effect of various parameters on the system performance.
\item Comparing the TSR and the PSR protocols, our numerical analysis shows that in delay-limited transmission mode, the throughput performance of the TSR protocol is superior to the PSR protocol at higher transmission rates, at relatively lower signal-to-noise-ratio (SNR), and for lower energy harvesting efficiency. This is in contrast with point-to-point system where the PS receiver architecture always achieve larger rate-energy pairs than the TS receiver architecture.
\item For both the TSR and the PSR protocols, our numerical results show that locating the relay node closer to the source node yields larger throughput in delay-limited and delay-tolerant transmission modes. This is also in contrast with the general case where energy harvesting is not considered at the relay and the maximum throughput occurs when the relay is located midway between the source and the destination.
\end{itemize}

\subsection{Related Works}

Some recent studies have considered energy harvesting through the RF signals in wireless cooperative networks \cite{Chalise-12-P,Fouladgar-12-A}. In \cite{Chalise-12-P}, the authors considered a MIMO relay system and studied different tradeoffs between the energy transfer and the information rates to achieve the optimal source and relay precoding. However, the authors in \cite{Chalise-12-P} assume that the relay has its own internal energy source and does not need external charging. In contrast to \cite{Chalise-12-P}, we consider the case that the relay relies on external charging through the RF signal from the source node. In \cite{Fouladgar-12-A}, the authors investigated multi-user and multi-hop systems for simultaneous information and power transfer. It was shown in \cite{Fouladgar-12-A} that for a dual-hop channel with an energy harvesting relay, the transmission strategy depends on the quality of the second link. However, in \cite{Fouladgar-12-A}, the optimization strategy assumed perfect channel state information at the transmitter, which is often not practical. Further, it is assumed in \cite{Fouladgar-12-A} that the relay node is able to decode information and extract power
simultaneously, which, as explained in \cite{Zhou-12-A}, may not hold in practice. In contrast to \cite{Fouladgar-12-A}, the system model proposed in this paper assumes the availability of the channel state information at the destination only and adopts the practical receiver architecture at the relay with separate information decoding and energy harvesting.

\subsection{Organization}

The remainder of the paper is organized as follows. Section \ref{sec:sys_mod} presents the overall system model and assumptions. Sections \ref{sec:TSA} and \ref{sec:PSA} detail the TSR and the PSR protocols, respectively, and analytically characterize the throughput performance. Section \ref{sec:ideal_rx} investigates the achievable throughput of an ideal receiver that is able to process information and extract power from the same received signal. Section \ref{sec:sim} presents the numerical results from which various design insights are obtained. Finally, Section \ref{sec:conclusions} concludes the paper and summarizes the key findings.

\section{System Model}\label{sec:sys_mod}
\begin{figure}[t]
\centering
     \hspace{-.0cm}\includegraphics[scale=1.3]{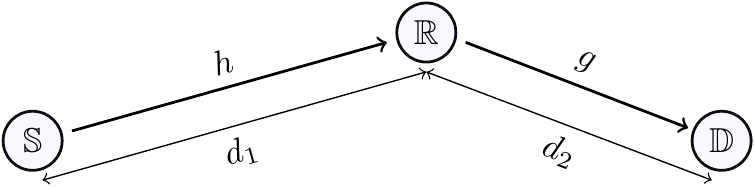}
     \centering
    \caption{System model for energy constrained relay assisted communication between a source and a destination node.}
    \label{fig:sys}
\end{figure}

A wireless communication system is considered, where the information is transferred from the source node, $\mathbb{S}$, to the destination node, $\mathbb{D}$, through an energy constrained intermediate relay node, $\mathbb{R}$. Fig. \ref{fig:sys} shows the system model for the considered system. The quasi-static block-fading channel gains from the source to the relay and from the relay to the destination nodes are denoted by $h$ and $g$, respectively. The distances from the source to the relay and from the relay to the destination nodes are denoted by $d_1$ and $d_2$, respectively. Throughout this paper, the following set of assumptions are considered.
\begin{enumerate}[{A}1.]
\item There is no direct link between the source and the destination node. Thus, an intermediate relay assists the transmission of the source messages to the destination \cite{Hasna-02-P}. A single relay node is considered for simplicity, as shown in Fig. \ref{fig:sys}.
\item The intermediate relay is an energy constrained node. It first harvests energy from the source signal. Then, it uses the harvested energy as a source of transmit power to forward the source information to the destination. It is assumed that the energy harvesting and information transfer are carried out for every received block without any constraint on the minimum power level of the received signal.
\item Amongst the different relaying protocols, amplify-and-forward (AF) scheme is chosen at the relay node due to its implementation simplicity \cite{Laneman-04-A}.
\item It is assumed that the processing power required by the transmit/receive circuitry at the relay is negligible as compared to the power used for signal transmission from the relay to the destination. This is justifiable when the transmission distances are large such that the energy transmitted is the dominant source of energy consumption \cite{Medepally-10-A,Zhou-12-A}.
\item The channel gains, $h$ and $g$ are modeled as quasi-static block-fading and frequency non-selective parameters. The channel is constant over the \emph{block time} $T$ and independent and identically distributed from one block to the next, following a Rayleigh distribution. The use of such channels is motivated by prior research in this field~\cite{Zhou-12-A,Zhang-12-A,Medepally-10-A,Ho-12-A,Liu-12-A,Luo-12-A}.
\item It is assumed that the channel state information is available only at the destination. The destination node can estimate the dual-hop channel at the start of information transmission in each block by utilizing the pilots sent from the source node over the dual-hop link. We have assumed perfect channel knowledge at the destination and assumed negligible overhead for pilot transmission, which is inline with the previous work in this research field \cite{Zhou-12-A,Zhang-12-A,Medepally-10-A,Ho-12-A,Liu-12-A,Luo-12-A}.

\end{enumerate}

Based on the time switching and the power splitting receiver architectures, we propose two relaying protocols to harvest energy from the source RF signal, i) TSR protocol and ii) PSR protocol. Moreover, we consider two different transmission modes, i) delay-limited and ii) delay-tolerant, which refer to the applicability of different length of the code-words \cite{Liu-12-A}. The delay-limited transmission mode implies that the destination node has to decode the received signal block by block and thus the code length cannot be larger than the transmission block time. On the other hand, the delay-tolerant transmission mode implies that the destination node can buffer the received information blocks and can tolerate the delay in decoding the received signal. Thus, the code length can be kept very large compared to the transmission block time. The detailed analysis of the proposed TSR and PSR protocols under both transmission modes is given in the following sections.

\section{Time Switching-Based Relaying (TSR) Protocol} \label{sec:TSA}

\ifCLASSOPTIONpeerreview

\else
 \newcounter{MYtempeqncnt}
 \begin{figure*}[!b]
 \hrulefill
 \vspace{-0.0cm}
\normalsize
\setcounter{MYtempeqncnt}{0}
\setcounter{equation}{9}
\begin{align}\label{eq:gD}
      \gamma_D &= \frac{ \frac{ 2 \eta  P_s^2 |h|^4 |g|^2 \alpha } { (1-\alpha) d_1^m d_2^m ( P_s |h|^2  + d_1^m \sigr ) } } {   \frac{ 2 \eta P_s |h|^2 |g|^2 \sigr \alpha } { (1-\alpha) d_2^m ( P_s |h|^2  + d_1^m \sigr ) } + \sigd } \notag \\
      & = \frac{2 \eta P_s^2 |h|^4 |g|^2 \alpha}{2 \eta P_s |h|^2 |g|^2 d_1^m \sigr \alpha + P_s |h|^2 d_1^m d_2^m \sigd (1- \alpha) + d_1^{2m} d_2^{m} \sigr \sigd (1-\alpha) }.
\end{align}
\setcounter{equation}{\value{MYtempeqncnt}}
\vspace*{-.4cm}
\end{figure*}
\fi

\begin{figure}[t]
  \centering
  \subfigure[]
  {
    \hspace*{-0.23in}
    \includegraphics[width=0.40 \textwidth]{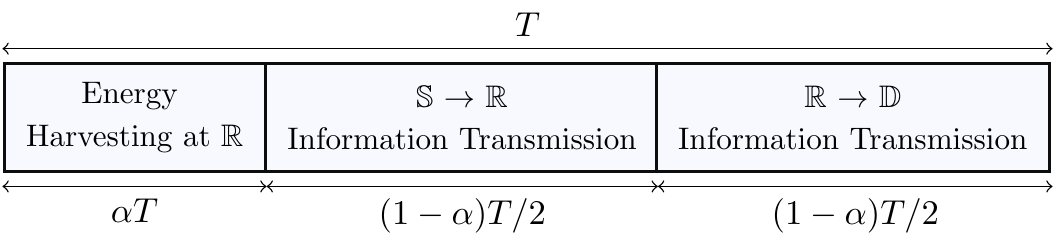}
    \label{fig:TS}
  }
  \subfigure[]
  {
    \includegraphics[width=0.50 \textwidth]{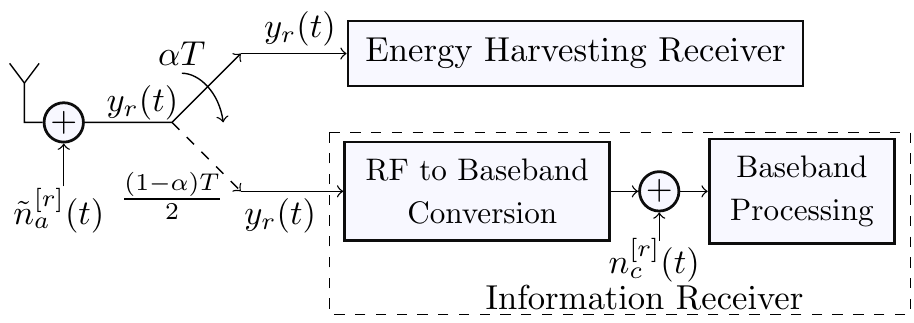}
    \label{fig:TS_R}
  }
 \caption{(a) Illustration of the key parameters in the TSR protocol for energy harvesting and information processing at the relay.  (b) Block diagram of the relay receiver in the TSR protocol.}
  \vspace{-0.11in}
\end{figure}

Fig. \ref{fig:TS} depicts the key parameters in the TSR protocol for energy harvesting and information processing at the relay. In Fig. \ref{fig:TS}, $T$ is the block time in which a certain block of information is transmitted from the source node to the destination node and $\alpha$ is the fraction of the block time in which relay harvests energy from the source signal, where $0 \le \alpha \le 1$. The remaining block time, $(1-\alpha) T$ is used for information transmission, such that half of that, $(1-\alpha) T/2$, is used for the source to relay information transmission and the remaining half, $(1-\alpha) T/2$, is used for the relay to destination information transmission. All the energy harvested during energy harvesting phase is consumed by the relay while forwarding the source signal to the destination. The choice of the time fraction, $\alpha$, used for harvesting energy at the relay node, affects the achievable throughput at the destination. The following subsections analyze the energy harvesting and information processing at the relay node.

\subsection{Energy Harvesting}

The block diagram for the relay receiver in the TSR protocol is shown in Fig. \ref{fig:TS_R}. The RF signal, $y(t)$ received at the relay node is first sent to the \emph{energy harvesting receiver} (for $\alpha T$ time) and then to the \emph{information receiver} (for $(1-\alpha) T/2$ time). Note that the RF signal, $y(t)$ is corrupted by the narrow-band Gaussian noise, $\tilde{n}_a^{[r]}(t)$, introduced by the receiving antenna.\footnote{Note that the superscript $[r]$, e.g., with the noise $\tilde{n}_a^{[r]}(t)$, is used to indicate the noise at the relay node.} The energy harvesting receiver rectifies the RF signal directly and gets the direct current to charge up the battery. The details of such an energy harvesting receiver can be found in \cite{Zhou-12-A}. As shown in Fig. \ref{fig:TS_R}, the received signal at the relay node, $y_r(t)$ is given by
\begin{align}\label{eq:yt}
      y_r(t) =  \frac{1}{\sqrt{d_1^m}} \sqrt{P_s} h s(t)  +  \tilde{n}_a^{[r]}(t),
\end{align}
where $h$ is the source to relay channel gain, $d_1$ is the source to relay distance, $P_s$ is the transmitted power from the source, $m$ is the path loss exponent, and $s(t)$ is the normalized information signal from the source, i.e., $\mathbb{E} \{ |s(t)|^2 \} = 1$, where $\mathbb{E} \{ \cdot \}$ is the expectation operator and $|\cdot|$ is the absolute value operator.

Using \eqref{eq:yt}, the harvested energy, $E_h$ during energy harvesting time $\alpha T$ is given by \cite{Zhou-12-A}\vspace{-5pt}
\begin{align}\label{eq:EH_TS}
      E_h =  \frac{\eta P_s |h|^2}{d_1^m} \alpha T,
\end{align}
where $0 < \eta < 1$ is the energy conversion efficiency which depends on the rectification process and the energy harvesting circuitry \cite{Zhou-12-A}.

\subsection{Energy Constrained Relay-Assisted Transmission}

The information receiver in Fig. \ref{fig:TS_R} down-converts the RF signal to baseband and processes the baseband signal, where $n_{c}^{[r]}(t)$ is the additive noise due to RF band to baseband signal conversion. After down conversion, the sampled baseband signal at the relay node, $y_r(k)$, is given by
\begin{align}\label{eq:rrk_AF}
      y_r(k) = \frac{1}{\sqrt{d_1^m}} \sqrt{P_s} h s(k)  +  n_a^{[r]}(k) + n_c^{[r]}(k),
\end{align}
where $k$ denotes the symbol index, $s(k)$ is the sampled and normalized information signal from the source, $n_a^{[r]}(k)$ is the baseband additive white Gaussian noise (AWGN) due to the receiving antenna at the relay node,\footnote{Note that $n_a^{[r]}(k)$ is the baseband equivalent noise of the pass band noise $\tilde{n}_a^{[r]}(t)$ \cite{Zhou-12-A}.} and $n_c^{[r]}(k)$ is the sampled AWGN due to RF band to baseband signal conversion. The relay amplifies the received signal and the transmitted signal from the relay, $x_r(k)$ is given by
\begin{align}\label{eq:trk_AF}
      x_r(k) =  \frac{\sqrt{P_r} y_r(k) } { \sqrt{ \frac{P_s |h|^2}{d_1^m} + \sigra + \sigrp } },
\end{align}
where the factor in the denominator, $\sqrt{ \frac{P_s |h|^2}{d_1^m} + \sigra + \sigrp }$ is the power constraint factor at the relay, $\sigra$ and $\sigrp$ are the variances of the AWGNs, $n_a^{[r]}(k)$ and $n_a^{[r]}(k)$, respectively, and $P_r$ is the power transmitted from the relay node, which depends on the amount of energy harvested during the energy harvesting time. Note that the relay node can obtain the power constraint factor from the power of the received signal. The sampled received signal at the destination, $y_d(k)$ is given by \vspace{-3pt}
\begin{align}\label{eq:rdk_AF}
      y_d(k) = \frac{1}{\sqrt{d_2^m}} g x_r(k)  +  n_a^{[d]}(k) + n_c^{[d]}(k),
\end{align}
where $n_a^{[d]}(k)$ and $n_c^{[d]}(k)$ are the antenna and conversion AWGNs at the destination node, respectively, and $g$ is the relay to destination channel gain. Substituting \eqref{eq:trk_AF} into \eqref{eq:rdk_AF}, we have
\ifCLASSOPTIONpeerreview
\begin{align}\label{eq:rdk2_AF}
      y_d(k) = \frac{g \sqrt{P_r d_1^m} y_r(k) } { \sqrt{d_2^m} \sqrt{ P_s |h|^2  + d_1^m (\sigra + \sigrp) } }  +  n_a^{[d]}(k) + n_c^{[d]}(k).
\end{align}
\else
\begin{align}\label{eq:rdk2_AF}
      y_d(k) &= \frac{g \sqrt{P_r d_1^m} y_r(k) } { \sqrt{d_2^m} \sqrt{ P_s |h|^2  + d_1^m (\sigra + \sigrp) } }  +  n_a^{[d]}(k) \notag \\ & \hspace{5.5cm} + n_c^{[d]}(k).
\end{align}
\fi

Finally, substituting $y_r(k)$ from \eqref{eq:rrk_AF} into \eqref{eq:rdk2_AF}, $y_d(k)$ is given by
\ifCLASSOPTIONpeerreview
\begin{align}\label{eq:rdk3_AF}
      y_d(k) = \frac{ \sqrt{P_r P_s} h g s(k)  } { \sqrt{d_2^m} \sqrt{ P_s |h|^2  + d_1^m \sigr } }  +  \frac{ \sqrt{P_r d_1^m} g n^{[r]}(k)  } { \sqrt{d_2^m} \sqrt{ P_s |h|^2  + d_1^m \sigr } } + n^{[d]}(k),
\end{align}
\else
\begin{align}\label{eq:rdk3_AF}
      y_d(k) &= \frac{ \sqrt{P_r P_s} h g s(k)  } { \sqrt{d_2^m} \sqrt{ P_s |h|^2  + d_1^m \sigr } }    \notag \\ & \hspace{1.5cm} +  \frac{ \sqrt{P_r d_1^m} g n^{[r]}(k)  } { \sqrt{d_2^m} \sqrt{ P_s |h|^2  +   d_1^m \sigr } } + n^{[d]}(k),
\end{align}
\fi
where $n^{[r]}(k) \triangleq n_a^{[r]}(k) + n_c^{[r]}(k)$ and $n^{[d]}(k) \triangleq n_a^{[d]}(k) + n_c^{[d]}(k)$ are the overall AWGNs at the relay and destination nodes, respectively and $ \sigr \triangleq \sigra + \sigrp $. Using $E_h$ in \eqref{eq:EH_TS}, the transmitted power from the relay node, $P_r$ is given by
\begin{align}\label{eq:Pr_TS}
      P_r = \frac{E_h}{(1- \alpha) T /2} = \frac{2 \eta P_s |h|^2 \alpha }{d_1^m (1 - \alpha) },
\end{align}
where \eqref{eq:Pr_TS} follows from the fact that relay communicates with the destination node for the time $(1- \alpha) T /2$, as shown in Fig. \ref{fig:TS}. Substituting the value of $P_r$ from \eqref{eq:Pr_TS} into \eqref{eq:rdk3_AF}, the received signal at the destination, $y_d(k)$ in terms of $P_s$, $\eta$, $\alpha$, $d_1$ and $d_2$, is given by
\ifCLASSOPTIONpeerreview
\begin{align}\label{eq:rdk4_AF}
      y_d(k) = \underbrace{\frac{ \sqrt{2 \eta |h|^2 \alpha} P_s h g s(k)  } { \sqrt{(1-\alpha) d_1^m d_2^m} \sqrt{ P_s |h|^2  + d_1^m \sigr } }}_\text{signal part}  +  \underbrace{\frac{ \sqrt{2 \eta P_s |h|^2 \alpha} g n^{[r]}(k)  } { \sqrt{(1-\alpha) d_2^m} \sqrt{ P_s |h|^2  + d_1^m \sigr } } + n^{[d]}(k)}_\text{overall noise}.
\end{align}
\else
\begin{align}\label{eq:rdk4_AF}
      y_d(k) &= \underbrace{\frac{ \sqrt{2 \eta |h|^2 \alpha} P_s h g s(k)  } { \sqrt{(1-\alpha) d_1^m d_2^m} \sqrt{ P_s |h|^2  + d_1^m \sigr } }}_\text{signal part} \notag \\  & \hspace{0.4cm} +  \underbrace{\frac{ \sqrt{2 \eta P_s |h|^2 \alpha} g n^{[r]}(k)  } { \sqrt{(1-\alpha) d_2^m} \sqrt{ P_s |h|^2  + d_1^m \sigr } } + n^{[d]}(k)}_\text{overall noise}.
\end{align}
\fi

\subsection{Throughput Analysis}

Using \eqref{eq:rdk4_AF}, the SNR at the destination node, $\gamma_D  =  \frac{ \mathbb{E} \{ |\text{signal part in \eqref{eq:rdk4_AF}}|^2 \}}{\mathbb{E} \{ |\text{overall noise in \eqref{eq:rdk4_AF}}|^2 \}} $  is given by
\ifCLASSOPTIONpeerreview
\begin{align}\label{eq:gD}
      \gamma_D &=   \frac{ \frac{ 2 \eta  P_s^2 |h|^4 |g|^2 \alpha } { (1-\alpha) d_1^m d_2^m ( P_s |h|^2  + d_1^m \sigr ) } } {   \frac{ 2 \eta P_s |h|^2 |g|^2 \sigr \alpha } { (1-\alpha) d_2^m ( P_s |h|^2  + d_1^m \sigr ) } + \sigd } \notag \\
      & = \frac{2 \eta P_s^2 |h|^4 |g|^2 \alpha}{2 \eta P_s |h|^2 |g|^2 d_1^m \sigr \alpha + P_s |h|^2 d_1^m d_2^m \sigd (1- \alpha) + d_1^{2m} d_2^{m} \sigr \sigd (1-\alpha) },
\end{align}
\else
\addtocounter{equation}{1}
\eqref{eq:gD} at the bottom of the page,
\fi
where $ \sigd \triangleq \sigda + \sigdp $. In the following, the throughput, $\tau$, is determined at the destination node, given the received SNR, $\gamma_D$ in \eqref{eq:gD}, for both the delay-limited and the delay-tolerant transmission modes.

\subsubsection{Delay-limited Transmission} \label{sec:AF_TS_SF}

In the delay-limited transmission mode, the throughput is determined by evaluating the outage probability, $p_\text{out}$, at a fixed source transmission rate, i.e., $R$ bits/sec/Hz, where $R \triangleq \log_2 ( 1 + \gamma_0 ) $ and $\gamma_0$ is the threshold value of SNR for correct data detection at the destination. Thus, $p_\text{out}$ is given by \ifCLASSOPTIONpeerreview \vspace{-15pt} \fi
\begin{align}\label{eq:pout}
      p_\text{out} = p( \gamma_D < \gamma_0 ),
\end{align}
where $ \gamma_0 = 2^R - 1$. The analytical expression for $\pout$ is given in the following proposition.

\emph{Proposition 1}: The outage probability at the destination node for the TSR protocol is given by
\begin{subequations}\label{eq:pout_A}
\ifCLASSOPTIONpeerreview
\begin{align}
      p_\text{out}  &=  1 - \displaystyle\frac{1}{\lambda_h} \int_{z = d/c}^{\infty}  e^ {- \left( \frac{z}{\lambda_h} + \frac{az+b}{(cz^2-dz)\lambda_g}   \right)  } dz  \label{eq:pout_A1} \\ &\approx 1 - e^{- \frac{d}{c \lambda_h}} u K_{1} \left(  u \right), \hspace{3cm} (\text{high SNR approximation}) \label{eq:pout_A2}
\end{align}
\else
\begin{align}
      p_\text{out}  &=  1 - \displaystyle\frac{1}{\lambda_h} \int_{z = d/c}^{\infty}  e^ {- \left( \frac{z}{\lambda_h} + \frac{az+b}{(cz^2-dz)\lambda_g}   \right)  } dz  \label{eq:pout_A1} \\ &\approx 1 - e^{- \frac{d}{c \lambda_h}} u K_{1} \left(  u \right), \hspace{0.8cm} (\text{high SNR approximation}) \label{eq:pout_A2}
\end{align}
\fi
\end{subequations}
where, \\ \vspace{-1cm}
\begin{subequations}\label{eq:Constants_1}
\begin{align}
a &\triangleq P_s d_1^m d_2^m \sigd \gamma_0 (1 - \alpha), \\
b &\triangleq d_1^{2m} d_2^m \sigr \sigd \gamma_0 (1 - \alpha),  \\
c &\triangleq 2 \eta P_s^2 \alpha, \\
d &\triangleq 2 \eta P_s d_1^m \sigr \gamma_0 \alpha, \\
u &\triangleq \sqrt{\frac{4a}{c \lh \lgg} },
\end{align}
\end{subequations}
$\lh$ and $\lgg$ are the mean values of the exponential random variables $|h|^2$ and $|g|^2$, respectively, and
$K_{1}(\cdot)$ is the first-order modified Bessel function of the second kind \cite{Gradshteyn-80-B}.

\emph{Proof}: See Appendix~\ref{app:A}.

Proposition 1 derives the outage probability at the destination when the relay harvests energy from the source signal and uses that harvested energy to forward the source signal to the destination. The outage probability, in \eqref{eq:pout_A}, is a function of the energy harvesting time $\alpha$ and decreases as $\alpha$ increases from $0$ to $1$. This is because a larger $\alpha$ results in more transmission power at the relay, which in turn decreases the chance of an outage. Given that the transmitter is communicating $R$ bits/sec/Hz and $(1 - \alpha) T /2$ is the effective communication time from the source node to the destination node in the block of time $T$ seconds, as shown in Fig. \ref{fig:TS}, the throughput, $\tau$ at the destination is given by
\begin{align}\label{eq:tau}
      \tau = (1 - \pout) R  \frac{(1 - \alpha) T /2 }{T} = \frac{(1 - \pout) R (1 - \alpha)}{2},
\end{align}
where the throughput, $\tau$ in \eqref{eq:tau}, depends on $P_s$, $\eta$, $\alpha$, $d_1$, $d_2$, $R$, $\sigr$ and $\sigd$.

\subsubsection{Delay-Tolerant Transmission}

In the delay-tolerant transmission mode, the throughput is determined by evaluating the ergodic capacity, $C$ at the destination. Unlike the delay-limited transmission mode, where the source transmits at fixed rate $R$ in order to meet some outage criteria, the source can transmit data at any rate less than or equal to the evaluated ergodic capacity, $C$ in the delay-tolerant transmission mode. In fact, the delay-tolerant transmission mode assumes that the code length is very large compared to the block time so that the code sees all the possible realizations of the channel during a code-word transmission and channel conditions average out. Thus, it is possible to achieve the ergodic capacity by transmitting at a rate equal to the ergodic capacity without any rate adaptation or requiring any knowledge about the channel state information at the source or the relay node \cite{Biglieri-98-A}. Using the received SNR at the destination, $\gamma_D$ in \eqref{eq:gD}, $C$ is given by\vspace{-2pt}
\begin{align}\label{eq:C1}
      C = \mathbb{E}_{h,g} \left\{ \log_2(1 + \gamma_D) \right\},
\end{align}
where $\gamma_D$ depends on the random channel gains, $h$ and $g$.

\emph{Proposition 2}: The ergodic capacity at the destination node for the TSR protocol is given by
\begin{subequations}\label{eq:C_A}
\ifCLASSOPTIONpeerreview
\begin{align}
      C  &=  \int_{\gamma = 0}^{\infty} \int_{z = d/c}^{\infty}  \frac{(az+b) c z^2}{(cz^2-dz)^2 \lambda_g \lambda_h \gamma} e^ {- \left( \frac{z}{\lambda_h} + \frac{az+b}{(cz^2-dz)\lambda_g}   \right)  }  \log_2(1 + \gamma)  dz d \gamma \label{eq:C_A1} \\ &\approx  \displaystyle\int_{\gamma = 0}^{\infty} \left( \frac{u^2 K_0(u) e^{- \frac{d}{c \lambda_h}} }{2 \gamma} + \frac{ d u K_{1}(u) e^{- \frac{d}{c \lambda_h}} }{ \gamma c \lambda_h}  \right) \log_2(1 + \gamma)  d \gamma, \hspace{1cm} (\text{high SNR approximation}) \label{eq:C_A2}
\end{align}
\else
\begin{align}
      C  &=  \int_{\gamma = 0}^{\infty} \int_{z = d/c}^{\infty}  \frac{(az+b) c z^2}{(cz^2-dz)^2 \lambda_g \lambda_h \gamma} e^ {- \left( \frac{z}{\lambda_h}  + \frac{az+b}{(cz^2-dz)\lambda_g}   \right)  } \notag \\ & \hspace{4.7cm} \log_2(1 + \gamma)  dz d \gamma \label{eq:C_A1} \\ &\approx  \displaystyle\int_{\gamma = 0}^{\infty} \left( \frac{u^2 K_0(u) e^{- \frac{d}{c \lambda_h}} }{2 \gamma} + \frac{ d u K_{1}(u) e^{- \frac{d}{c \lambda_h}} }{ \gamma c \lambda_h}  \right) \notag \\ & \hspace{1.7cm} \log_2(1 + \gamma)  d \gamma, \hspace{0.5cm} (\text{high SNR approximation}) \label{eq:C_A2}
\end{align}
\fi
\end{subequations}
where,
\begin{subequations}\label{eq:Constants_2}
\begin{align}
a &\triangleq P_s d_1^m d_2^m \sigd \gamma (1 - \alpha), \\
b &\triangleq d_1^{2m} d_2^m \sigr \sigd \gamma (1 - \alpha),  \\
c &\triangleq 2 \eta P_s^2 \alpha, \\
d &\triangleq 2 \eta P_s d_1^m \sigr \gamma \alpha, \\
u &\triangleq \sqrt{\frac{4a}{c \lh \lgg} },
\end{align}
\end{subequations}
and $\lh$ and $\lgg$ are defined below \eqref{eq:pout_A}.

\emph{Proof}: See Appendix~\ref{app:B}.

Proposition 2 derives the ergodic capacity at the destination when the relay harvests energy from the source signal and uses that harvested energy to forward the source signal to the destination. The ergodic capacity, in \eqref{eq:C_A}, is a function of the energy harvesting time $\alpha$ and increases as $\alpha$ increases from $0$ to $1$. This is because a larger $\alpha$ results in more transmission power at the relay, which in turn increases the ergodic capacity. On the other hand, the effective communication time between the source node and the destination node, $(1 - \alpha) T /2$, decreases by increasing $\alpha$. Thus, throughput $\tau$ is not an increasing function of $\alpha$. Given that the source is transmitting at a fixed rate equal to the ergodic capacity, i.e., $C$ bits/sec/Hz, the throughput, $\tau$ at the destination is given by
\begin{align}\label{eq:tau_C}
      \tau = \frac{(1 - \alpha) T /2 }{T} C = \frac{(1 - \alpha) }{2} C ,
\end{align}
where the throughput, $\tau$ in \eqref{eq:tau_C} depends on $P_s$, $\eta$, $\alpha$, $d_1$, $d_2$, $\sigr$ and $\sigd$. Note that the final throughput expressions in the delay-limited and the delay-tolerant transmission modes, in \eqref{eq:tau} and \eqref{eq:tau_C}, respectively, also take into account the energy harvesting time, $\alpha T$ and depend only on the effective information transmission time, $(1 - \alpha) T /2$.

\section{Power Splitting-Based Relaying (PSR) Protocol} \label{sec:PSA}

\begin{figure}[t]
  \centering
  \subfigure[]
  {
    \hspace*{-0.23in}
    \includegraphics[width=0.40 \textwidth]{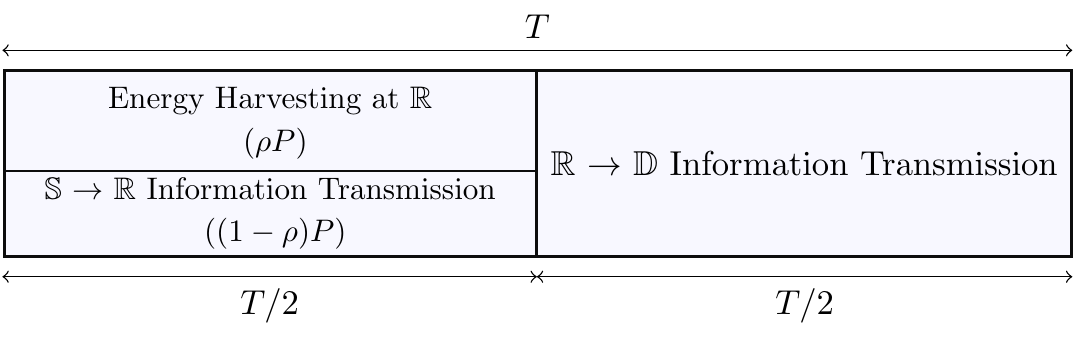}
    \label{fig:PS}
  }
  \subfigure[]
  {
    \includegraphics[width=0.50 \textwidth]{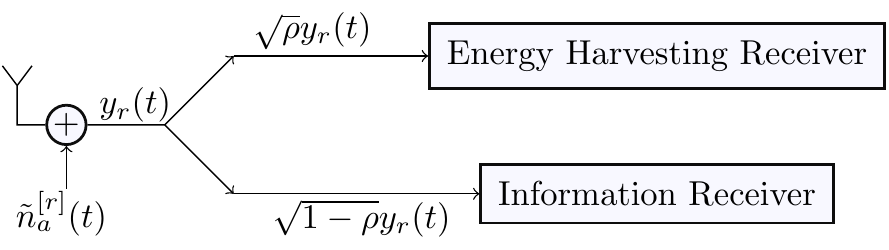}
    \label{fig:PS_R}
  }
  \caption{(a) Illustration of the key parameters in the PSR protocol for energy harvesting and information processing at the relay.  (b) Block diagram of the relay receiver in the PSR protocol (the details of information receiver are the same as shown in Fig. \ref{fig:TS_R}).}
  \vspace{-0.11in}
\end{figure}

Fig. \ref{fig:PS} shows the communication block diagram employing the PSR protocol for energy harvesting and information processing at the relay. In Fig. \ref{fig:PS}, $P$ is the power of the received signal, $y_r(t)$ at the relay and $T$ is the total block time, from which half of the time, $T/2$ is used for the source to relay information transmission and the remaining half, $T/2$ is used for the relay to destination information transmission. During the first half of the block time, the fraction of the received signal power, $\rho P$ is used for energy harvesting and the remaining received power, $ (1-\rho)P $ is used for the source to relay information transmission, where $0 \le \rho \le 1$. All the harvested energy is consumed by the relay while forwarding the source signal to the destination. The choice of the power fraction, $\rho$, used for harvesting energy at the relay node, affects the achievable throughput at the destination. The following subsections analyze the energy harvesting and information processing at the relay for the PSR protocol.

\subsection{Energy Harvesting}

The block diagram for the relay receiver in the PSR protocol is shown in Fig. \ref{fig:PS_R}. The power splitter splits the received signal in $\rho:1-\rho$ proportion, such that the portion of the received signal, $\sqrt{\rho} y_r(t)$ is sent to the energy harvesting receiver and the remaining signal strength, $\sqrt{1-\rho} y_r(t)$ drives the information receiver. Using the signal received at the input of the energy harvesting receiver, $\sqrt{\rho} y_r(t) =  \frac{1}{\sqrt{d_1^m}} \sqrt{\rho P_s} h s(t)  + \sqrt{\rho}  \tilde{n}_a^{[r]}(t) $, the harvested energy, $E_h$ at the relay is given by \cite{Zhou-12-A} \vspace{-0pt}
\begin{align}\label{eq:EH_PS}
      E_h =  \frac{\eta \rho P_s |h|^2}{d_1^m} (T/2),
\end{align}
where the energy is harvested at the relay during half of the block time, $T/2$, as shown in Fig. \ref{fig:PS}, and $0 < \eta < 1$ is the energy conversion efficiency.

\subsection{Energy Constrained Relay-Assisted Transmission}

\ifCLASSOPTIONpeerreview
\else
\begin{figure*}[!b]
 \hrulefill
 \vspace{-0.0cm}
\normalsize
\setcounter{MYtempeqncnt}{19}
\setcounter{equation}{25}
\begin{align}\label{eq:gD_PS}
      \gamma_D = \displaystyle\frac{\eta  P_s^2 |h|^4 |g|^2 \rho (1-\rho)}{ \eta P_s |h|^2 |g|^2 d_1^m \sigr \rho + P_s |h|^2 d_1^m d_2^m \sigd (1- \rho) + d_1^{2m} d_2^{m} \sigr \sigd  }
\end{align}
\setcounter{equation}{\value{MYtempeqncnt}}
\vspace*{-.4cm}
\end{figure*}
\fi

The block level description of the information receiver in Fig. \ref{fig:PS_R} is the same as that detailed in Fig. \ref{fig:TS_R}. After down conversion, the sampled baseband signal, $y_r(k)$, at the input of basedband processor in the PSR protocol is given by
\begin{align}\label{eq:rrk_AF_PS}
      y_r(k) = \frac{1}{\sqrt{d_1^m}} \sqrt{(1-\rho)P_s} h s(k)  + \sqrt{(1-\rho)} n_a^{[r]}(k) + n_c^{[r]}(k),
\end{align}
where $s(k)$, $h$, $P_s$, $n_a^{[r]}(k)$, and $n_c^{[r]}(k)$ are defined below \eqref{eq:rrk_AF} and $\rho$ is the portion of the received power used for energy harvesting, as explained at the start of Section \ref{sec:PSA}. The relay amplifies the received signal and the transmitted signal from the relay is given by
\begin{align}\label{eq:trk_AF_PS}
      x_r(k) =  \frac{\sqrt{P_r} y_r(k) } { \sqrt{(1-\rho) \frac{P_s |h|^2 }{d_1^m} + (1-\rho) \sigra + \sigrp } },
\end{align}
where the factor in the denominator, $\sqrt{(1-\rho) \frac{P_s |h|^2 }{d_1^m} + (1-\rho) \sigra + \sigrp }$ is the power constraint factor at the relay, $P_r$ is the power transmitted from the relay and $\sigra$ and $\sigrp$ are defined below \eqref{eq:trk_AF}. Substituting \eqref{eq:trk_AF_PS} into \eqref{eq:rdk_AF}, the sampled received signal at the destination node, $y_d(k)$ in the PSR protocol is given by
\ifCLASSOPTIONpeerreview
\begin{align}\label{eq:rdk2_AF_PS}
      y_d(k) = \frac{g \sqrt{P_r d_1^m} y_r(k) } { \sqrt{d_2^m} \sqrt{ (1-\rho) P_s |h|^2  + d_1^m ((1-\rho)\sigra + \sigrp) } }  +  n_a^{[d]}(k) + n_c^{[d]}(k).
\end{align}
\else
\begin{align}\label{eq:rdk2_AF_PS}
      y_d(k) &= \frac{g \sqrt{P_r d_1^m} y_r(k) } { \sqrt{d_2^m} \sqrt{ (1-\rho) P_s |h|^2  + d_1^m ((1-\rho)\sigra + \sigrp) } } \notag \\ & \hspace{4.0cm} +  n_a^{[d]}(k) + n_c^{[d]}(k).
\end{align}
\fi

Finally, substituting \eqref{eq:rrk_AF_PS} into \eqref{eq:rdk2_AF_PS}, $y_d(k)$ is given by
\ifCLASSOPTIONpeerreview
\begin{align}\label{eq:rdk3_AF_PS}
      y_d(k) = \frac{ \sqrt{(1-\rho) P_s P_r} h g s(k)  } { \sqrt{d_2^m} \sqrt{ (1-\rho) P_s |h|^2  + d_1^m \sigr } }  +  \frac{ \sqrt{P_r d_1^m} g n^{[r]}(k)  } { \sqrt{d_2^m} \sqrt{ (1-\rho) P_s |h|^2  + d_1^m \sigr } } + n^{[d]}(k),
\end{align}
\else
\begin{align}\label{eq:rdk3_AF_PS}
      y_d(k) &= \frac{ \sqrt{(1-\rho) P_s P_r} h g s(k)  } { \sqrt{d_2^m} \sqrt{ (1-\rho) P_s |h|^2  + d_1^m \sigr } } \notag \\ & \hspace{0.2cm} +  \frac{ \sqrt{P_r d_1^m} g n^{[r]}(k)  } { \sqrt{d_2^m} \sqrt{ (1-\rho) P_s |h|^2  + d_1^m \sigr } } + n^{[d]}(k),
\end{align}
\fi
where $n^{[r]}(k) \triangleq \sqrt{1-\rho} n_a^{[r]}(k) + n_c^{[r]}(k)$ and $n^{[d]}(k) \triangleq n_a^{[d]}(k) + n_c^{[d]}(k)$ are the overall AWGNs at the relay and the destination nodes, respectively and $ \sigr \triangleq (1-\rho) \sigra + \sigrp $. Note that the definitions of $n^{[r]}(k)$ and $ \sigr$ in the PSR protocol are different from the TSR protocol. Using $E_h$ in \eqref{eq:EH_PS}, the transmitted power from the relay node, $P_r$ is given by
\begin{align}\label{eq:Pr_PS}
      P_r = \frac{E_h}{T /2} = \frac{\eta P_s |h|^2 \rho }{d_1^m},
\end{align}
where \eqref{eq:Pr_PS} follows from the fact that the relay communicates with the destination node for half of the block time $T /2$, as shown in Fig. \ref{fig:PS}. Substituting the value of $P_r$ from \eqref{eq:Pr_PS} into \eqref{eq:rdk3_AF_PS}, the received signal at the destination, $y_d(k)$ in terms of $P_s$, $\eta$, $\rho$, $d_1$ and $d_2$, is given by
\ifCLASSOPTIONpeerreview
\begin{align}\label{eq:rdk4_AF_PS}
      y_d(k) = \underbrace{\frac{ \sqrt{\eta |h|^2 \rho (1-\rho)} P_s h g s(k)  } { \sqrt{d_1^m d_2^m} \sqrt{ P_s |h|^2 (1-\rho) + d_1^m \sigr } }}_\text{signal part}  +  \underbrace{\frac{ \sqrt{\eta P_s |h|^2 \rho} g n^{[r]}(k)  } { \sqrt{d_2^m} \sqrt{ P_s |h|^2 (1-\rho) + d_1^m \sigr } } + n^{[d]}(k)}_\text{overall noise}.
\end{align}
\else
\begin{align}\label{eq:rdk4_AF_PS}
      y_d(k) &= \underbrace{\frac{ \sqrt{\eta |h|^2 \rho (1-\rho)} P_s h g s(k)  } { \sqrt{d_1^m d_2^m} \sqrt{ P_s |h|^2 (1-\rho) + d_1^m \sigr } }}_\text{signal part}  \notag \\ & +  \underbrace{\frac{ \sqrt{\eta P_s |h|^2 \rho} g n^{[r]}(k)  } { \sqrt{d_2^m} \sqrt{ P_s |h|^2 (1-\rho) + d_1^m \sigr } } + n^{[d]}(k)}_\text{overall noise}.
\end{align}
\fi

\subsection{Throughput Analysis}

Using \eqref{eq:rdk4_AF_PS}, the SNR at the destination node, $\gamma_D =  \frac{ \mathbb{E} \{ |\text{signal part in \eqref{eq:rdk4_AF_PS}}|^2 \}}{\mathbb{E} \{ |\text{overall noise in \eqref{eq:rdk4_AF_PS}}|^2 \}} $ in case of the PSR protocol is given by
\ifCLASSOPTIONpeerreview
\vspace{-14pt}
\begin{align}\label{eq:gD_PS}
      \gamma_D = \displaystyle\frac{\eta  P_s^2 |h|^4 |g|^2 \rho (1-\rho)}{ \eta P_s |h|^2 |g|^2 d_1^m \sigr \rho + P_s |h|^2 d_1^m d_2^m \sigd (1- \rho) + d_1^{2m} d_2^{m} \sigr \sigd  }
\end{align}
\else
\addtocounter{equation}{1}
\eqref{eq:gD_PS} at the bottom of the page,
\fi
where $ \sigd \triangleq \sigda + \sigdp $. In the following, we determine the throughput, $\tau$, at the destination node for the PSR protocol, given the received SNR, $\gamma_D$ in \eqref{eq:gD_PS}, for both the delay-limited and the delay-tolerant transmission modes.

\subsubsection{Delay-Limited Transmission}

Given that the transmitter is communicating $R$ bits/sec/Hz and $T /2$ is the effective communication time from the source node to the destination node in the block of time $T$ seconds, as shown in Fig. \ref{fig:PS}, the throughput, $\tau$ at the destination node in the delay-limited transmission mode is given by\vspace{-10pt}
\begin{align}\label{eq:tau_PS}
      \tau =  (1-\pout) R  \frac{ T/2} {T} =  \frac{(1 - \pout) R}{2}
\end{align}
where the outage probability, $\pout$ can be calculated using Proposition 3 (see below) for $\gamma_D$ given in \eqref{eq:gD_PS} and $\gamma_0$ defined below \eqref{eq:pout}.

\emph{Proposition 3}: For the PSR protocol, $\pout$ can be analytically calculated using \eqref{eq:pout_A}, where\footnote{The detailed derivation of $\pout$ for the PSR protocol is omitted here because it follows the same steps as given in Appendix \ref{app:A}.}
\begin{subequations}\label{eq:Constants_3}
\begin{align}
a &\triangleq P_s d_1^m d_2^m \sigd \gamma_0 (1 - \rho), \\
b &\triangleq d_1^{2m} d_2^m \sigr \sigd \gamma_0,  \\
c &\triangleq \eta P_s^2 \rho (1 - \rho), \\
d &\triangleq \eta P_s d_1^m \sigr \gamma_0 \rho , \hspace{0.4cm} \text{and}\\
u &\triangleq \sqrt{\frac{4a}{c \lh \lgg} }.
\end{align}
\end{subequations}

The throughput, $\tau$ in \eqref{eq:tau_PS} depends on $P_s$, $\eta$, $\rho$, $d_1$, $d_2$, $R$, $\sigr$ and $\sigd$. The expression for the outage probability, $\pout$ in \eqref{eq:pout_A}, seems similar for both the TSR and the PSR protocols. However, this is not the case because the final expressions for $\pout$ are written in terms of constants $a$, $b$, $c$, and $d$, which differ in the TSR and the PSR protocols

\subsubsection{Delay-Tolerant Transmission}

Since $T /2$ is the effective communication time between the source and the destination nodes in the block of time $T$ seconds, as shown in Fig. \ref{fig:PS}, the throughput, $\tau$ at the destination node in the delay-tolerant transmission mode is given by
\begin{align}\label{eq:tau_C_PS}
      \tau = C \frac{T /2 }{T} = \frac{C}{2},
\end{align}
where the ergodic capacity, $C$ can be calculated using Proposition 4 (see below) for $\gamma_D$ given in \eqref{eq:gD_PS}.

\emph{Proposition 4}: For the PSR protocol, $C$ can be analytically calculated using \eqref{eq:C_A}, where\footnote{The detailed derivation of $C$ for the PSR protocol is omitted here because it follows the same steps as given in Appendix \ref{app:B}.}
\begin{subequations}\label{eq:Constants_4}
\begin{align}
a &\triangleq P_s d_1^m d_2^m \sigd \gamma (1 - \rho), \\
b &\triangleq d_1^{2m} d_2^m \sigr \sigd \gamma,  \\
c &\triangleq \eta P_s^2 \rho (1 - \rho), \\
d &\triangleq \eta P_s d_1^m \sigr \gamma \rho , \hspace{0.4cm} \text{and}\\
u &\triangleq \sqrt{\frac{4a}{c \lh \lgg} }.
\end{align}
\end{subequations}

The derived formulae for throughput for both the TSR and the PSR protocols are summarized in Table \ref{tab:AF}. The ergodic capacity, $C$ in \eqref{eq:C_A}, has been expressed in terms of the constants $a$, $b$, $c$, and $d$, which differ in the TSR and PSR protocols and are defined in Table \ref{tab:AF}.

\ifCLASSOPTIONpeerreview

\begin{table}[t]
\vspace{1cm}
\caption{Throughput ($\tau$) for the TSR and PSR protocols with Amplify and Forward Relaying.} \centering
\begin{tabular}{|c||c|c|} \hline
 \cline{2-3} \multirow{3}{*}{} & TSR Protocol & PSR Protocol \\ \cline{2-3} %
& $\gamma_D = \displaystyle\frac{2 \eta P_s^2 |h|^4 |g|^2 \alpha}{2 \eta P_s |h|^2 |g|^2 d_1^m \sigr \alpha + P_s |h|^2 d_1^m d_2^m \sigd (1- \alpha)  }$ & $\gamma_D = \displaystyle\frac{\eta  P_s^2 |h|^4 |g|^2 \rho (1-\rho)}{ \eta P_s |h|^2 |g|^2 d_1^m \sigr \rho + P_s |h|^2 d_1^m d_2^m \sigd (1- \rho)   }$ \vspace{-0.1cm} \\
     &  \hspace{4cm} $+ d_1^{2m} d_2^{m} \sigr \sigd (1-\alpha)$               & \hspace{4.5cm}   $+ d_1^{2m} d_2^{m} \sigr \sigd$              \\
     & $\sigr = \sigra + \sigrp$ & $\sigr = (1 - \rho) \sigra + \sigrp$ \\
     & $\sigd = \sigda + \sigdp$ & $\sigr = \sigda + \sigdp$ \\
     & $ u = \sqrt{\frac{4a}{c \lh \lgg} } $ & $ u = \sqrt{\frac{4a}{c \lh \lgg} } $ \\ \hline \hline \multirow{8}{*}{\begin{sideways}Delay-Limited Transmission Mode\end{sideways}}   &
\multicolumn{2}{|c|}{$ p_\text{out}  =  1 - \displaystyle\frac{1}{\lambda_h} \displaystyle\int_{z = d/c}^{\infty}  e^ {- \left( \frac{z}{\lambda_h} + \frac{az+b}{(cz^2-dz)\lambda_g}   \right)  } dz$  (Analytical)}  \\  &
\multicolumn{2}{|c|}{$ p_\text{out}  \approx  1 - e^{- \frac{d}{c \lambda_h}} u K_{1} \left(  u \right)   $ (Analytical Approximation)}   \\ \cline{2-3}
&  $\tau =  (1-p_\text{out})(1 - \alpha) R / 2  $ &  $\tau =  (1-p_\text{out}) R / 2  $       \\ & $a = P_s d_1^m d_2^m \sigd \gamma_0 (1 - \alpha) $ & $a = P_s d_1^m d_2^m \sigd \gamma_0 (1 - \rho) $ \\
     & $b = d_1^{2m} d_2^m \sigr \sigd \gamma_0 (1 - \alpha) $ & $b  = d_1^{2m} d_2^m \sigr \sigd \gamma_0 $ \\
     & $c = 2 \eta P_s^2 \alpha $ & $c = \eta P_s^2 \rho (1 - \rho)$ \\
     & $d = 2 \eta P_s d_1^m \sigr \gamma_0 \alpha $ & $d = \eta P_s d_1^m \sigr \gamma_0 \rho $ \\ \hline \hline \multirow{9}{*}{\begin{sideways}Delay-Tolerant Transmission Mode\end{sideways}}   &
\multicolumn{2}{|c|}{$ C =  \displaystyle\int_{\gamma = 0}^{\infty} \int_{z = d/c}^{\infty}  \frac{(az+b) c z^2}{(cz^2-dz)^2 \lambda_g \lambda_h \gamma} e^ {- \left( \frac{z}{\lambda_h} + \frac{az+b}{(cz^2-dz)\lambda_g}   \right)  }  \log_2(1 + \gamma)  dz d \gamma  $  (Analytical)}  \\  &
\multicolumn{2}{|c|}{$ C  \approx  \displaystyle\int_{\gamma = 0}^{\infty} \left( \frac{u^2 K_0(u) e^{- \frac{d}{c \lambda_h}} }{2 \gamma} + \frac{ d u K_{1}(u) e^{- \frac{d}{c \lambda_h}} }{ \gamma c \lambda_h}  \right) \log_2(1 + \gamma)  d \gamma  $ (Analytical Approximation)}   \\ \cline{2-3}
& $\tau =  (1 - \alpha) C / 2  $        &   $\tau = C/2$      \\
     & $a = P_s d_1^m d_2^m \sigd \gamma_ (1 - \alpha) $ & $a = P_s d_1^m d_2^m \sigd \gamma (1 - \rho) $ \\
     & $b = d_1^{2m} d_2^m \sigr \sigd \gamma (1 - \alpha) $ & $b  = d_1^{2m} d_2^m \sigr \sigd \gamma $ \\
     & $c = 2 \eta P_s^2 \alpha $ & $c = \eta P_s^2 \rho (1 - \rho)$ \\
     & $d = 2 \eta P_s d_1^m \sigr \gamma \alpha $ & $d = \eta P_s d_1^m \sigr \gamma \rho $ \\ \hline
\end{tabular}
\label{tab:AF}
\end{table}

\else

\begin{table*}[t]
\vspace{1cm}
\caption{Throughput ($\tau$) for the TSR and PSR protocols with Amplify and Forward Relaying.} \centering
\begin{tabular}{|c||c|c|} \hline
 \cline{2-3} \multirow{3}{*}{} & TSR Protocol & PSR Protocol \\ \cline{2-3} %
& $\gamma_D = \displaystyle\frac{2 \eta P_s^2 |h|^4 |g|^2 \alpha}{2 \eta P_s |h|^2 |g|^2 d_1^m \sigr \alpha + P_s |h|^2 d_1^m d_2^m \sigd (1- \alpha)  }$ & $\gamma_D = \displaystyle\frac{\eta  P_s^2 |h|^4 |g|^2 \rho (1-\rho)}{ \eta P_s |h|^2 |g|^2 d_1^m \sigr \rho + P_s |h|^2 d_1^m d_2^m \sigd (1- \rho)   }$  \\
     &  \hspace{4cm} $+ d_1^{2m} d_2^{m} \sigr \sigd (1-\alpha)$               & \hspace{4.5cm}   $+ d_1^{2m} d_2^{m} \sigr \sigd$              \\
     & $\sigr = \sigra + \sigrp$ & $\sigr = (1 - \rho) \sigra + \sigrp$ \\
     & $\sigd = \sigda + \sigdp$ & $\sigr = \sigda + \sigdp$ \\
     & $ u = \sqrt{\frac{4a}{c \lh \lgg} } $ & $ u = \sqrt{\frac{4a}{c \lh \lgg} } $ \\ \hline \hline \multirow{8}{*}{\begin{sideways}Delay-Limited Transmission Mode\end{sideways}}   &
\multicolumn{2}{|c|}{$ p_\text{out}  =  1 - \displaystyle\frac{1}{\lambda_h} \displaystyle\int_{z = d/c}^{\infty}  e^ {- \left( \frac{z}{\lambda_h} + \frac{az+b}{(cz^2-dz)\lambda_g}   \right)  } dz$  (Analytical)}  \\  &
\multicolumn{2}{|c|}{$ p_\text{out}  \approx  1 - e^{- \frac{d}{c \lambda_h}} u K_{1} \left(  u \right)   $ (Analytical Approximation)}   \\ \cline{2-3}
&  & \\ & & \\ & $\tau =  (1-p_\text{out})(1 - \alpha) R / 2  $ &  $\tau =  (1-p_\text{out}) R / 2  $       \\ & $a = P_s d_1^m d_2^m \sigd \gamma_0 (1 - \alpha) $ & $a = P_s d_1^m d_2^m \sigd \gamma_0 (1 - \rho) $ \\
     & $b = d_1^{2m} d_2^m \sigr \sigd \gamma_0 (1 - \alpha) $ & $b  = d_1^{2m} d_2^m \sigr \sigd \gamma_0 $ \\
     & $c = 2 \eta P_s^2 \alpha $ & $c = \eta P_s^2 \rho (1 - \rho)$ \\
     & $d = 2 \eta P_s d_1^m \sigr \gamma_0 \alpha $ & $d = \eta P_s d_1^m \sigr \gamma_0 \rho $ \\ & & \\ & & \\ \hline \hline \multirow{9}{*}{\begin{sideways}Delay-Tolerant Transmission Mode\end{sideways}}   &
\multicolumn{2}{|c|}{$ C =  \displaystyle\int_{\gamma = 0}^{\infty} \int_{z = d/c}^{\infty}  \frac{(az+b) c z^2}{(cz^2-dz)^2 \lambda_g \lambda_h \gamma} e^ {- \left( \frac{z}{\lambda_h} + \frac{az+b}{(cz^2-dz)\lambda_g}   \right)  }  \log_2(1 + \gamma)  dz d \gamma  $  (Analytical)}  \\  &
\multicolumn{2}{|c|}{$ C  \approx  \displaystyle\int_{\gamma = 0}^{\infty} \left( \frac{u^2 K_0(u) e^{- \frac{d}{c \lambda_h}} }{2 \gamma} + \frac{ d u K_{1}(u) e^{- \frac{d}{c \lambda_h}} }{ \gamma c \lambda_h}  \right) \log_2(1 + \gamma)  d \gamma  $ (Analytical Approximation)}   \\ \cline{2-3}
& & \\ & & \\ & $\tau =  (1 - \alpha) C / 2  $        &   $\tau = C/2$      \\
     & $a = P_s d_1^m d_2^m \sigd \gamma_ (1 - \alpha) $ & $a = P_s d_1^m d_2^m \sigd \gamma (1 - \rho) $ \\
     & $b = d_1^{2m} d_2^m \sigr \sigd \gamma (1 - \alpha) $ & $b  = d_1^{2m} d_2^m \sigr \sigd \gamma $ \\
     & $c = 2 \eta P_s^2 \alpha $ & $c = \eta P_s^2 \rho (1 - \rho)$ \\
     & $d = 2 \eta P_s d_1^m \sigr \gamma \alpha $ & $d = \eta P_s d_1^m \sigr \gamma \rho $ \\ & & \\ & & \\ \hline
\end{tabular}
\label{tab:AF}
\end{table*}

\fi

\newtheorem{remark}{Remark}
\begin{remark}\label{rem:1}
       \emph{As shown in Table \ref{tab:AF}, the final expressions for the throughput, $\tau$ at the destination node depend either on the outage probability, $\pout$, (delay-limited transmission) or the ergodic capacity, $C$, (delay-tolerant transmission), which in turn depend on energy harvesting time, $\alpha$ for the TSR protocol and power splitting factor, $\rho$ for the PSR protocol.}

        \emph{It is desirable to find the values of $\alpha$ and $\rho$, that result in the maximum value of throughput, $\tau$, for the TSR and the PSR protcols, respectively. Because of the integrations and the Bessel functions involved in the analytical expressions of $\pout$ and $C$, as shown in Table \ref{tab:AF}, it seems intractable to evaluate the closed-form expressions for the optimal value of $\alpha$ and $\rho$ in terms of $\tau$. However, the optimization can be done offline by numerically evaluating the optimal values of $\alpha$ and $\rho$ for the given system parameters, including, source power $P_s$, energy harvesting efficiency $\eta$, source to relay distance $d_1$, relay to destination distance $d_2$, source transmission rate $R$, and noise variances $\sigr$ and $\sigd$.}
\end{remark}
\begin{remark}\label{rem:2}
\emph{The throughput expressions derived in this paper represent the upper bound on the practically achievable throughput. This is because the system constraints, such as propagation delay, finite alphabet modulation, a minimum power level required for energy harvesting, and automatic repeat requests or retransmissions in the case of packet loss affect the achievable system throughput.}
\end{remark}

\section{Ideal Relay Receiver} \label{sec:ideal_rx}

In this section, we analyze the throughput performance with an ideal relay receiver, which processes the information and extracts power from the same received signal \cite{Zhou-12-A}. Thus, during the first half of the block time, $T/2$, the relay node harvests energy and processes the information from the source signal and during the remaining block time, $T/2$, the relay node uses the harvested energy to forward the source signal to the destination. The harvested energy during the energy harvesting time $T/2$ is given by $E_h = \frac{\eta P_s |h|^2}{d_1^m} (T/2) $. Using this harvested energy, the transmitted power from the relay node, $P_r$ is given by
\begin{align}\label{eq:Pr_ID}
      P_r = \frac{E_h}{T /2} = \frac{\eta P_s |h|^2 }{d_1^m },
\end{align}
The general expression for the received signal at the destination node, $y_d(k)$, is given by \eqref{eq:rdk3_AF}. Substituting the value of $P_r$ from \eqref{eq:Pr_ID} into \eqref{eq:rdk3_AF}, $y_d(k)$ for the ideal receiver is given by
\ifCLASSOPTIONpeerreview
\begin{align}\label{eq:rdk4_AF_ID}
      y_d(k) = \underbrace{\frac{ \sqrt{ \eta |h|^2 } P_s h g s(k)  } { \sqrt{ d_1^m d_2^m} \sqrt{ P_s |h|^2  + d_1^m \sigr } }}_\text{signal part}  +  \underbrace{\frac{ \sqrt{\eta P_s |h|^2 } g n^{[r]}(k)  } { \sqrt{ d_2^m} \sqrt{ P_s |h|^2  + d_1^m \sigr } } + n^{[d]}(k)}_\text{overall noise}.
\end{align}
\else
\begin{align}\label{eq:rdk4_AF_ID}
      y_d(k) &= \underbrace{\frac{ \sqrt{ \eta |h|^2 } P_s h g s(k)  } { \sqrt{ d_1^m d_2^m} \sqrt{ P_s |h|^2  + d_1^m \sigr } }}_\text{signal part} \notag \\  & \hspace{0.4cm} +  \underbrace{\frac{ \sqrt{\eta P_s |h|^2 } g n^{[r]}(k)  } { \sqrt{ d_2^m} \sqrt{ P_s |h|^2  + d_1^m \sigr } } + n^{[d]}(k)}_\text{overall noise}.
\end{align}
\fi
Using \eqref{eq:rdk4_AF_ID}, the SNR at the destination node, $\gamma_D  =  \frac{ \mathbb{E} \{ |\text{signal part in \eqref{eq:rdk4_AF_ID}}|^2 \}}{\mathbb{E} \{ |\text{overall noise in \eqref{eq:rdk4_AF_ID}}|^2 \}} $  is given by
\ifCLASSOPTIONpeerreview
\begin{align}\label{eq:gD_ID}
      \gamma_D = \frac{\eta P_s^2 |h|^4 |g|^2 }{\eta P_s |h|^2 |g|^2 d_1^m \sigr + P_s |h|^2 d_1^m d_2^m \sigd + d_1^{2m} d_2^{m} \sigr \sigd },
\end{align}
\else
\begin{align}\label{eq:gD_ID}
      \gamma_D = \frac{\eta P_s^2 |h|^4 |g|^2 }{\eta P_s |h|^2 |g|^2 d_1^m \sigr +  d_1^m d_2^m \sigd (P_s |h|^2 + d_1^{m} \sigr) },
\end{align}
\fi
In the following, we determine the throughput, $\tau$, at the destination node for the ideal receiver, given the received SNR, $\gamma_D$ in \eqref{eq:gD_ID}, for both the delay-limited and the delay-tolerant transmission modes.\footnote{The detailed derivation for the ideal receiver is omitted here as it follows the same steps as given in Appendices \ref{app:A} and \ref{app:B}.}

\ifCLASSOPTIONpeerreview

\else
\begin{figure*}[t]
  \centering
  \vspace{-.8cm}
  \subfigure[]
  {
    \hspace*{-0.28in}
    \includegraphics[width=0.48 \textwidth]{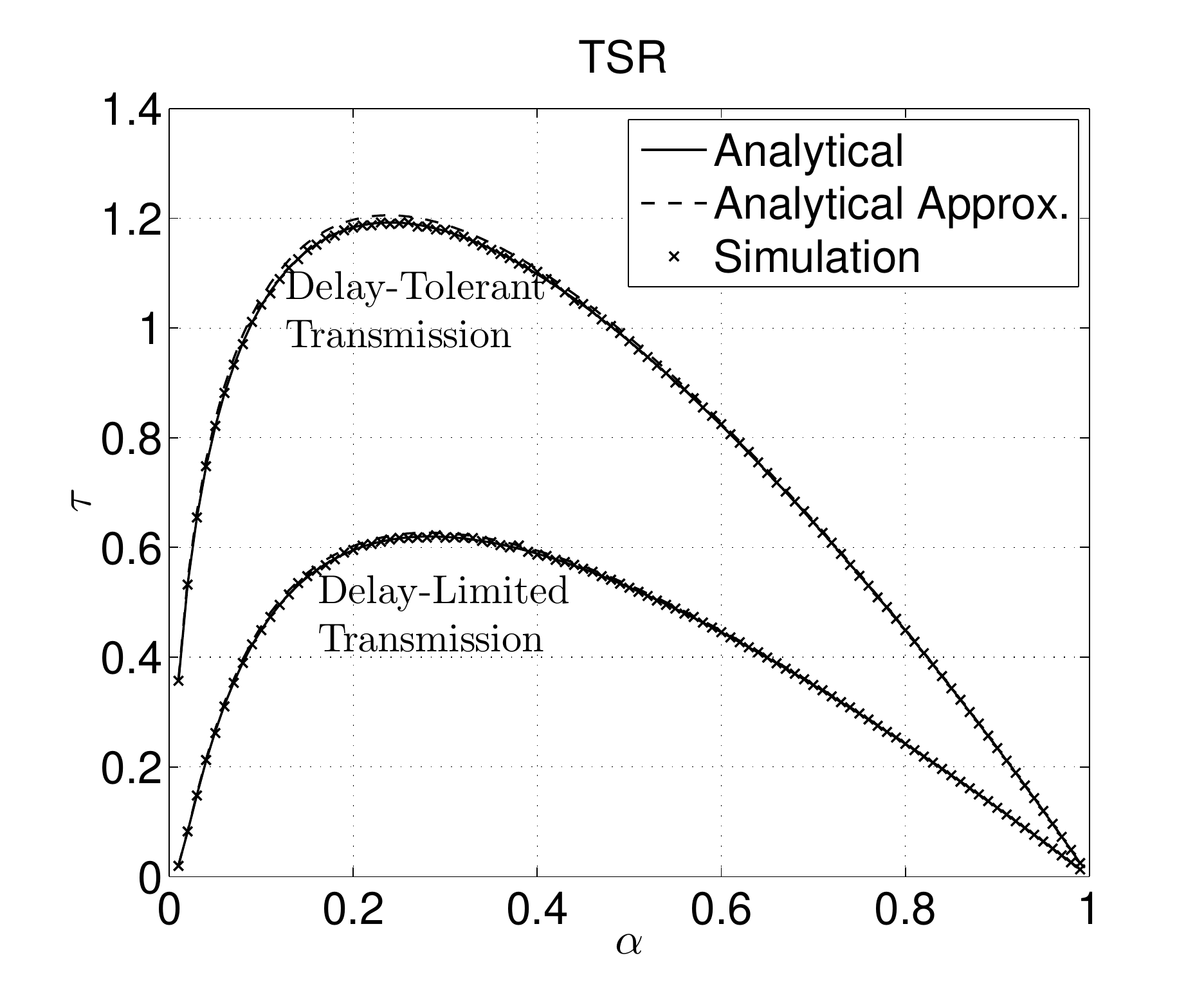}
    \label{fig:SF_TS_alpha}
  }
  \subfigure[]
  {
    \hspace{-0.0in}
    \includegraphics[width=0.48 \textwidth]{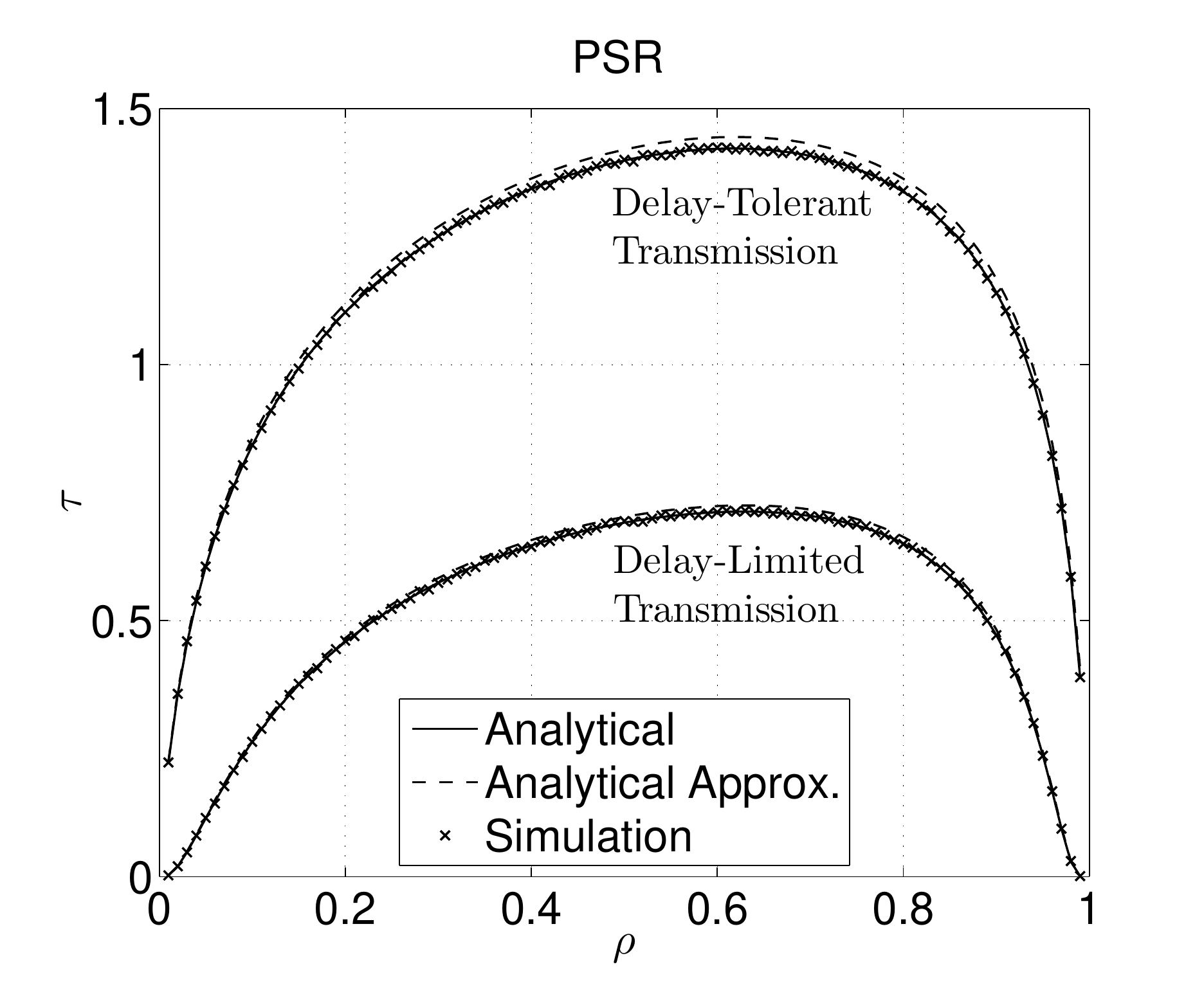}
    \label{fig:SF_PS_alpha}
  }
 \vspace{-0.05in}
  \caption{ Throughput $\tau$ at the destination node with respect to (a) $\alpha$ for the TSR protocol and (b) $\rho $ for the PSR protocol. Other parameters: $\sigma_{n_a}^2 = \sigma_{n_c}^2 = 0.01$, $P_s = 1$, $\eta = 1$, and $d_1 = d_2 = 1$.}
  \vspace{-0.10in}
\label{fig:SF_alpha}
\end{figure*}
\fi

\textit{1) Delay-Limited Transmission:} The transmission rate is $R$ bits/sec/Hz and the effective communication time between the source and the destination is $T /2$ for an ideal receiver. Thus, the throughput, $\tau$ at the destination node is given by $\tau = \frac{(1-\pout) R}{2}$, where the outage probability $\pout$ can be analytically calculated using \eqref{eq:pout_A} for $a \triangleq P_s d_1^m d_2^m \sigd \gamma_0$, $b \triangleq d_1^{2m} d_2^m \sigr \sigd \gamma_0$, $c \triangleq \eta P_s^2$, and $d \triangleq \eta P_s d_1^m \sigr \gamma_0$.

\textit{2) Delay-Tolerant Transmission:} The transmission rate is $C$ bits/sec/Hz. Using the effective communication time ($T/2$) between the source and the destination, the throughput, $\tau$ at the destination node is given by $\tau = \frac{C}{2}$, where the ergodic capacity $C$ can be analytically calculated using \eqref{eq:C_A} for $a \triangleq P_s d_1^m d_2^m \sigd \gamma$, $b \triangleq d_1^{2m} d_2^m \sigr \sigd \gamma$, $c \triangleq \eta P_s^2$, and $d \triangleq \eta P_s d_1^m \sigr \gamma$.

\section{Numerical Results and Discussion}\label{sec:sim}

This section uses the derived analytical results to provide insights into the various design choices. The optimal value of throughput $\tau$, optimal value of energy harvesting time $\alpha$ in the TSR protocol, and optimal value of power splitting ratio $\rho$ in the PSR protocol are investigated for different values of the noise variances, the source to relay and the relay to destination distances, $d_1$ and $d_2$, respectively, source transmission rate, $R$ and energy harvesting efficiency, $\eta$. \emph{The optimal values values of $\alpha$ and $\rho$ are numerically obtained as explained in Remark 1. Note that the optimal values of $\alpha$ and $\rho$ are defined as the values, which result in the maximum throughput $\tau$ at the destination node.}

Unless otherwise stated, we set the source transmission rate, $R = 3$ bits/sec/Hz in the delay limited transmission mode, energy harvesting efficiency, $\eta = 1$, source transmission power, $P_s = 1$ Joules/sec and path loss exponent $m = 2.7$ (which corresponds to an urban cellular network environment \cite{Meyr-98-B}). The distances $d_1$ and $d_2$ are normalized to unit value. For simplicity, similar noise variances at the relay and the destination nodes are assumed, i.e., antenna noise variance, $\siga \triangleq \sigra = \sigda$ and conversion noise variance, $\sigp \triangleq \sigra = \sigdp$. The mean values, $\lh$ and $\lgg$, of the exponential random variables $|h|^2$ and $|g|^2$, respectively, are set to $1$.

\subsection{Verification of Analytical Results}

In this subsection, the analytical results for the throughput, $\tau$, as shown in Table \ref{tab:AF}, are examined and verified through simulations for both the TSR and the PSR protocols in the delay-limited and the delay-tolerant transmission modes. Note that in order to calculate $\tau$, the analytical results for $\pout$ and $C$ are evaluated using \eqref{eq:pout_A} and \eqref{eq:C_A}, respectively and the simulation results for $\pout$ and $C$ are obtained using \eqref{eq:pout} and \eqref{eq:C1}, respectively. The simulation results in \eqref{eq:pout} and \eqref{eq:C1} are obtained by averaging these expressions over $10^5$ random realizations of the Rayleigh fading channels $h$ and $g$.

\ifCLASSOPTIONpeerreview

\begin{figure*}[t]
  \centering
  \vspace{-.8cm}
  \subfigure[]
  {
    \hspace*{-0.28in}
    \includegraphics[width=0.48 \textwidth]{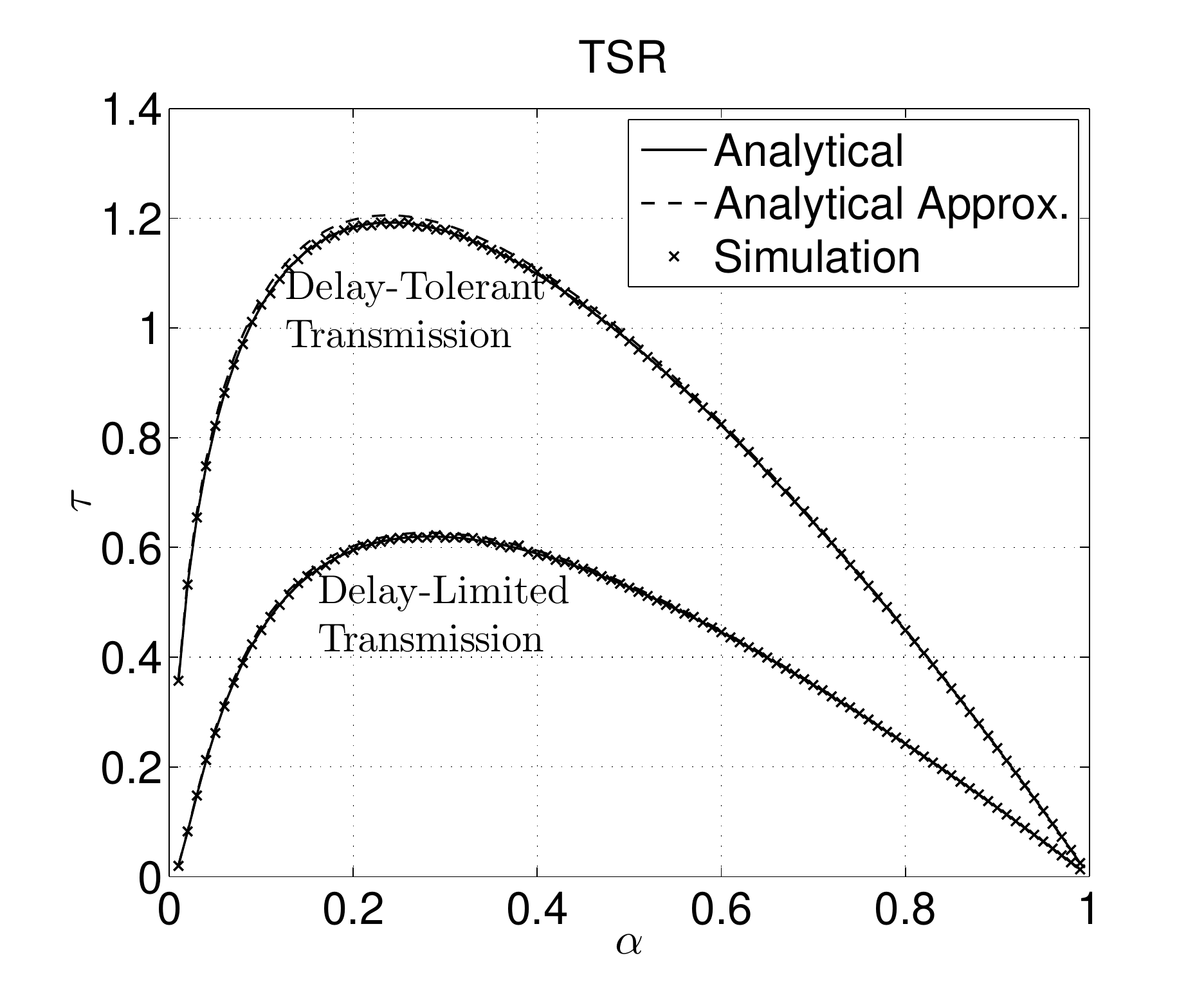}
    \label{fig:SF_TS_alpha}
  }
  \subfigure[]
  {
    \hspace{-0.0in}
    \includegraphics[width=0.48 \textwidth]{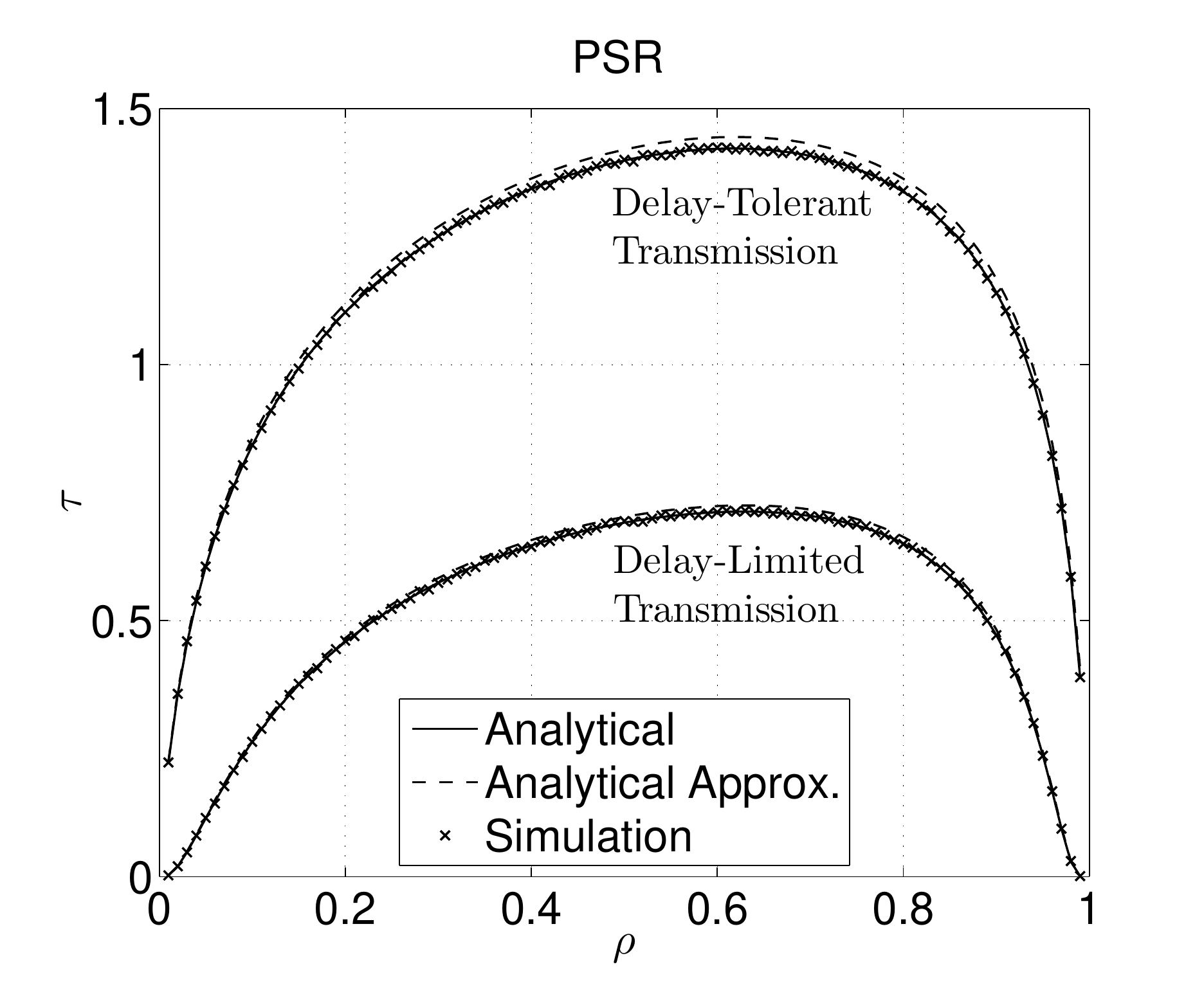}
    \label{fig:SF_PS_alpha}
  }
 \vspace{-0.05in}
  \caption{ Throughput $\tau$ at the destination node with respect to (a) $\alpha$ for the TSR protocol and (b) $\rho $ for the PSR protocol. Other parameters: $\sigma_{n_a}^2 = \sigma_{n_c}^2 = 0.01$, $P_s = 1$, $\eta = 1$, and $d_1 = d_2 = 1$.}
  \vspace{-0.10in}
\label{fig:SF_alpha}
\end{figure*}
\else
\fi

Fig. \ref{fig:SF_alpha} plots the throughput $\tau$ with respect to $0 < \alpha < 1$ for the TSR protocol (see Fig. \ref{fig:SF_TS_alpha}) and $0 < \rho < 1$ for the PSR protocol (see Fig. \ref{fig:SF_PS_alpha}). Both the antenna noise variance, $\sigma_{n_a}^2$ and the conversion noise variance, $\sigma_{n_c}^2$ are set to $0.01$. In order to evaluate the throughput for the delay-limited and the delay-tolerant transmission modes, the derived analytical and the analytical approximation (defined in figure as ``Analytical Approx.") expressions for the outage probability, $\pout$, and the ergodic capacity, $C$, as summarized in Table \ref{tab:AF}, are used. It can be observed from Fig. \ref{fig:SF_alpha} that the analytical and the simulation results match for all possible values of $\alpha$ and $\rho$ for both the TSR and the PSR protocols. This verifies the analytical expression for $\pout$ and $C$ presented in \emph{Proposition 1} and \emph{Proposition 2}, respectively.\footnote{The agreement between the analytical and simulation results is also observed for other values of antenna noise variance, $\sigma_{n_a}^2$ or conversion noise variance, $\sigma_{n_c}^2$. However, for brevity, the results for only one set of values of $\sigma_{n_a}^2$ and $\sigma_{n_c}^2$ are plotted in Fig. \ref{fig:SF_alpha}.} Fig. \ref{fig:SF_alpha} shows that the achievable throughput in the delay-tolerant transmission mode is more than in the delay-limited transmission mode for both the TSR and the PSR protocols. Fig. \ref{fig:SF_alpha} also shows that the closed-form analytical approximation results are very close to the exact analytical results.

\ifCLASSOPTIONpeerreview
\else
\begin{figure*}[t]
  \centering
  \vspace{-.8cm}
  \subfigure[]
  {
    \hspace*{-0.28in}
    \includegraphics[width=0.48 \textwidth]{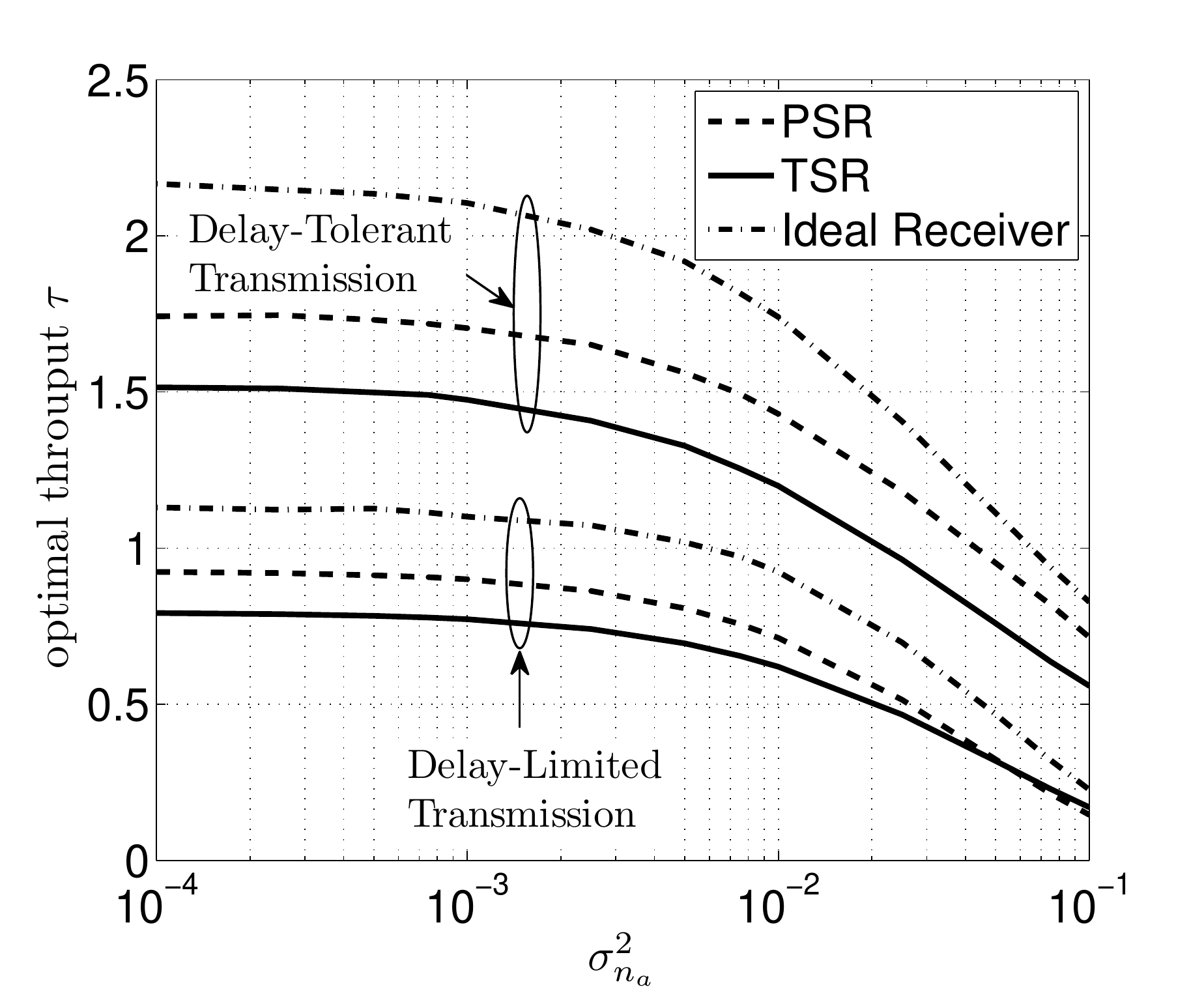}
    \label{fig:SF_noise_a}
  }
  \subfigure[]
  {
    \hspace{-0.0in}
    \includegraphics[width=0.48 \textwidth]{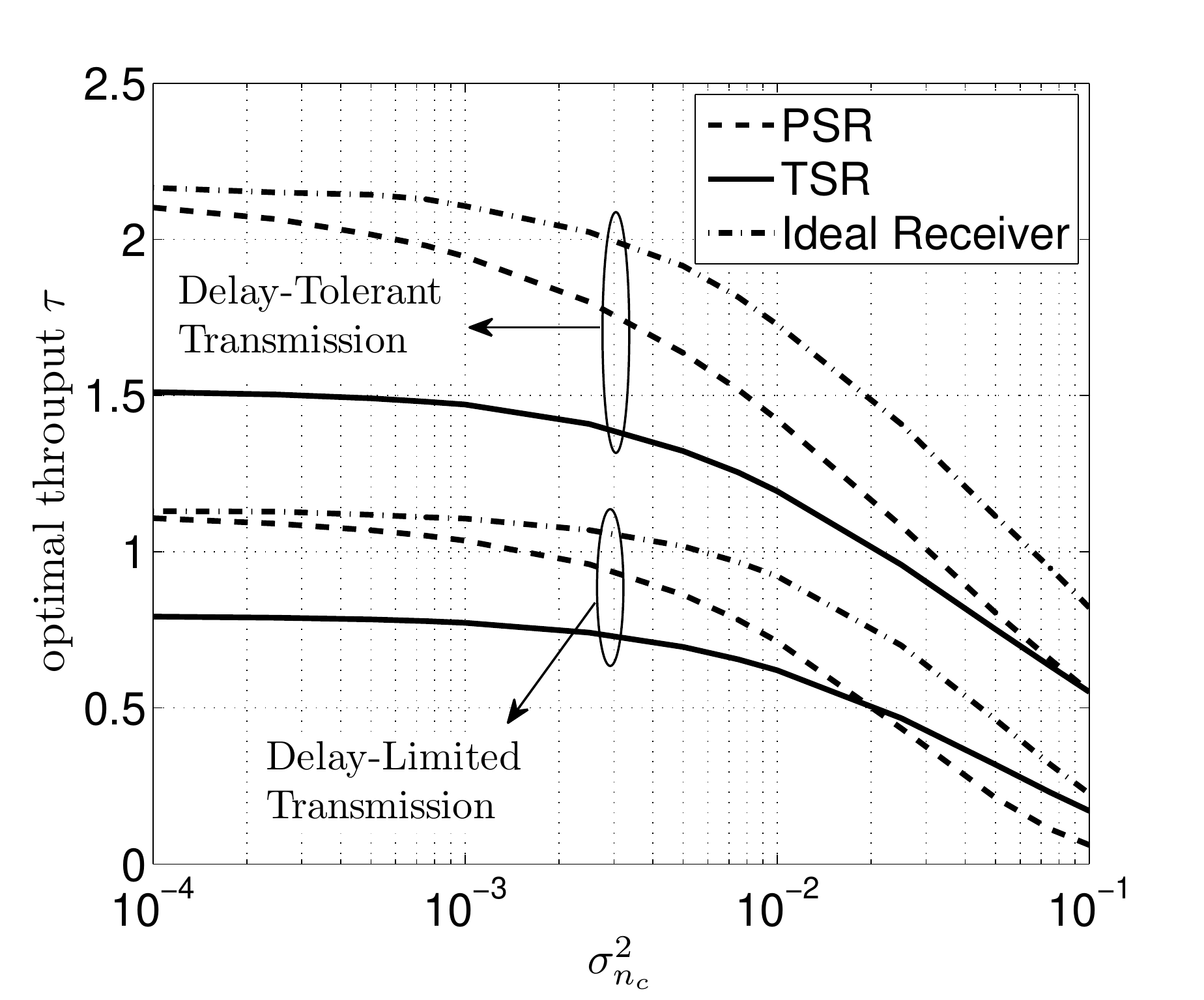}
    \label{fig:SF_noise_p}
  }
 \vspace{-0.05in}
  \caption{Optimal throughput $\tau$ for the ideal receiver, the TSR and the PSR protocols for (a) different values of antenna noise variance $\sigma_{n_a}^2$ and $\sigma_{n_c}^2 = 0.01$ (fixed) and (b) different values of conversion noise variance $\sigma_{n_c}^2$ and $\sigma_{n_a}^2 = 0.01$ (fixed). Other parameters: $P_s = 1$, $\eta = 1$, and $d_1 = d_2 = 1$.}
  \vspace{-0.10in}
\label{fig:SF_noise}
\end{figure*}

\fi
\subsection{Effect of Energy Harvesting Time, $\alpha$ (TSR protocol) and Power Splitting Factor, $\rho$ (PSR protocol)}

Fig. \ref{fig:SF_TS_alpha} shows that for the delay-limited transmission mode in the TSR protocol, the throughput increases as $\alpha$ increases from $0$ to some optimal $\alpha$ ($0.28$ for $\siga = 0.01$) but later, it starts decreasing as $\alpha$ increases from its optimal value. This is because for the values of $\alpha$ smaller than the optimal $\alpha$, there is less time for energy harvesting. Consequently, less energy is harvested and smaller values of throughput are observed at the destination node due to larger outage probability (see \eqref{eq:tau}). On the other hand, for the values of $\alpha$ greater than the optimal $\alpha$, more time is wasted on energy harvesting and less time is available for information transmission. As a result, smaller throughput results at the destination node due to smaller value of $(1-\alpha)/2$ (see \eqref{eq:tau}). A similar trend is observed for the delay-tolerant transmission mode in the TSR protocol.

Fig. \ref{fig:SF_PS_alpha} shows that for the delay-limited transmission mode in the PSR protocol, the throughput increases as $\rho$ increases from $0$ to some optimal $\rho$ ($0.63$ for $\siga = 0.01$) but later, it starts decreasing as $\rho$ increases from its optimal value. This is because for the values of $\rho$ smaller than the optimal $\rho$, there is less power available for energy harvesting. Consequently, less transmission power $P_r$ is available from the relay node and smaller values of throughput are observed at the destination node due to larger outage probability (see \eqref{eq:tau_PS}). On the other hand, for the values of $\rho$ greater than the optimal $\rho$, more power is wasted on energy harvesting and less power is left for the source to relay information transmission. As a result, poor signal strength is observed at the relay node and when the relay amplifies and forwards that noisy signal to the destination, larger outage occurs and results in lesser throughput at the destination node. A similar trend is observed for the delay-tolerant transmission mode in the PSR protocol.

\subsection{Effect of Noise Power}

Fig. \ref{fig:SF_noise} plots the optimal throughput $\tau$ for the ideal receiver, the TSR and the PSR protocols for different values of antenna noise variance, $\sigma_{n_a}^2$ (see Fig. \ref{fig:SF_noise_a} for fixed $\sigma_{n_c}^2 = 0.01$) and different values of conversion noise variance, $\sigma_{n_c}^2$ (see Fig. \ref{fig:SF_noise_p} for fixed $\sigma_{n_a}^2 = 0.01$).
\ifCLASSOPTIONpeerreview
\begin{figure*}[t]
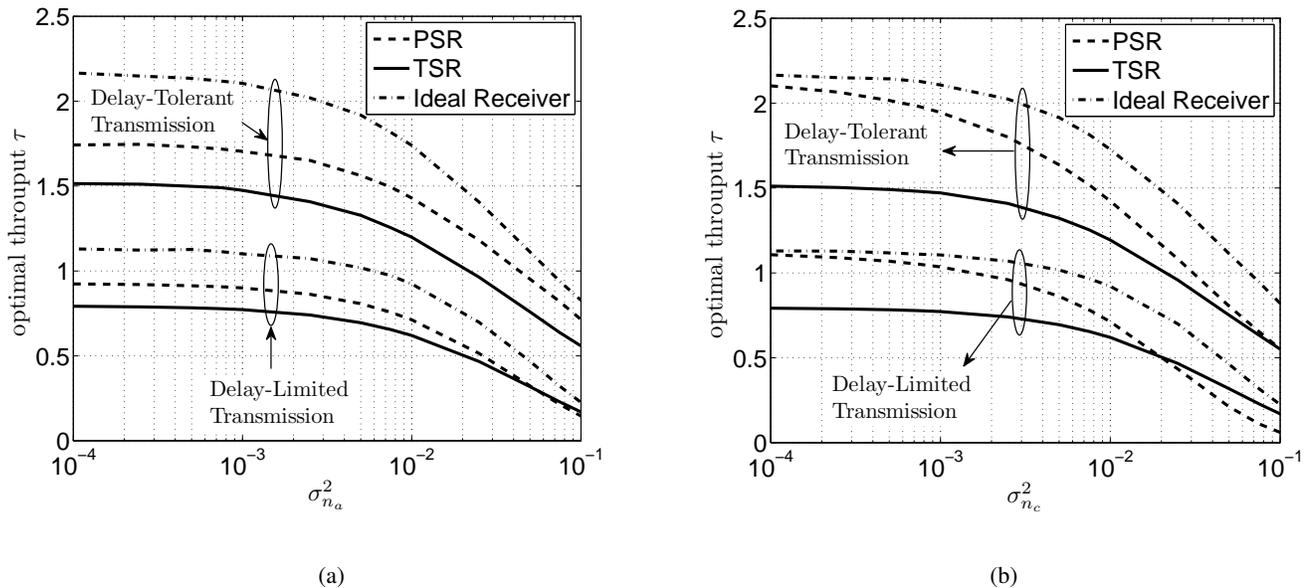

  \centering
  \vspace{-.8cm}
  \subfigure[]
  {
    \hspace*{-0.28in}
    \includegraphics[width=0.48 \textwidth]{noise_a_id}
    \label{fig:SF_noise_a}
  }
  \subfigure[]
  {
    \hspace{-0.0in}
    \includegraphics[width=0.48 \textwidth]{noise_p_id}
    \label{fig:SF_noise_p}
  }
 \vspace{-0.05in}
  \caption{Optimal throughput $\tau$ for the ideal receiver, the TSR and the PSR protocols for (a) different values of antenna noise variance $\sigma_{n_a}^2$ and $\sigma_{n_c}^2 = 0.01$ (fixed) and (b) different values of conversion noise variance $\sigma_{n_c}^2$ and $\sigma_{n_a}^2 = 0.01$ (fixed). Other parameters: $P_s = 1$, $\eta = 1$, and $d_1 = d_2 = 1$.}
  \vspace{-0.10in}
\label{fig:SF_noise}
\end{figure*}
\else
\fi
The throughput expressions for the ideal receiver are derived in Section \ref{sec:ideal_rx}. Since, the ideal receiver is based on the assumption of processing information and extracting power from the same received signal, the throughput performance of the ideal receiver outperforms the proposed TSR and PSR protocols for different values of the noise variances, as shown in Fig. \ref{fig:SF_noise}. It is interesting to note that the throughput performance gap between the TSR protocol and the ideal receiver becomes constant, as noise variances, $\siga$ or $\sigp$, approach $0$, On the other hand, Fig. \ref{fig:SF_noise_p} shows that the throughput performance gap between the PSR protocol and the ideal receiver decreases, as conversion noise variance $\sigp$ approaches $0$. \emph{Comparing the TSR and the PSR protocols for the delay-limited transmission mode, Fig. \ref{fig:SF_noise} illustrates that the PSR protocol is better than the TSR protocol to obtain larger values of the throughput, except at relatively large noise variance, where the TSR protocol achieves more throughput than the PSR protocol.} The crossover between the performances of the PSR and TSR protocols occurs at $\siga = 0.06$ (Fig. \ref{fig:SF_noise_a}) and $\sigp = 0.02$ (Fig. \ref{fig:SF_noise_p}). On the other hand, it can be observed from Fig. \ref{fig:SF_noise} that in the case of the delay-tolerant transmission mode, the PSR protocol is superior to the TSR protocol for the considered values of the noise variance to obtain larger values of the throughput.

\begin{figure*}[t]
  \centering
  \vspace{-.0cm}
  \subfigure[]
  {
    \hspace*{-0.28in}
    \includegraphics[width=0.48 \textwidth]{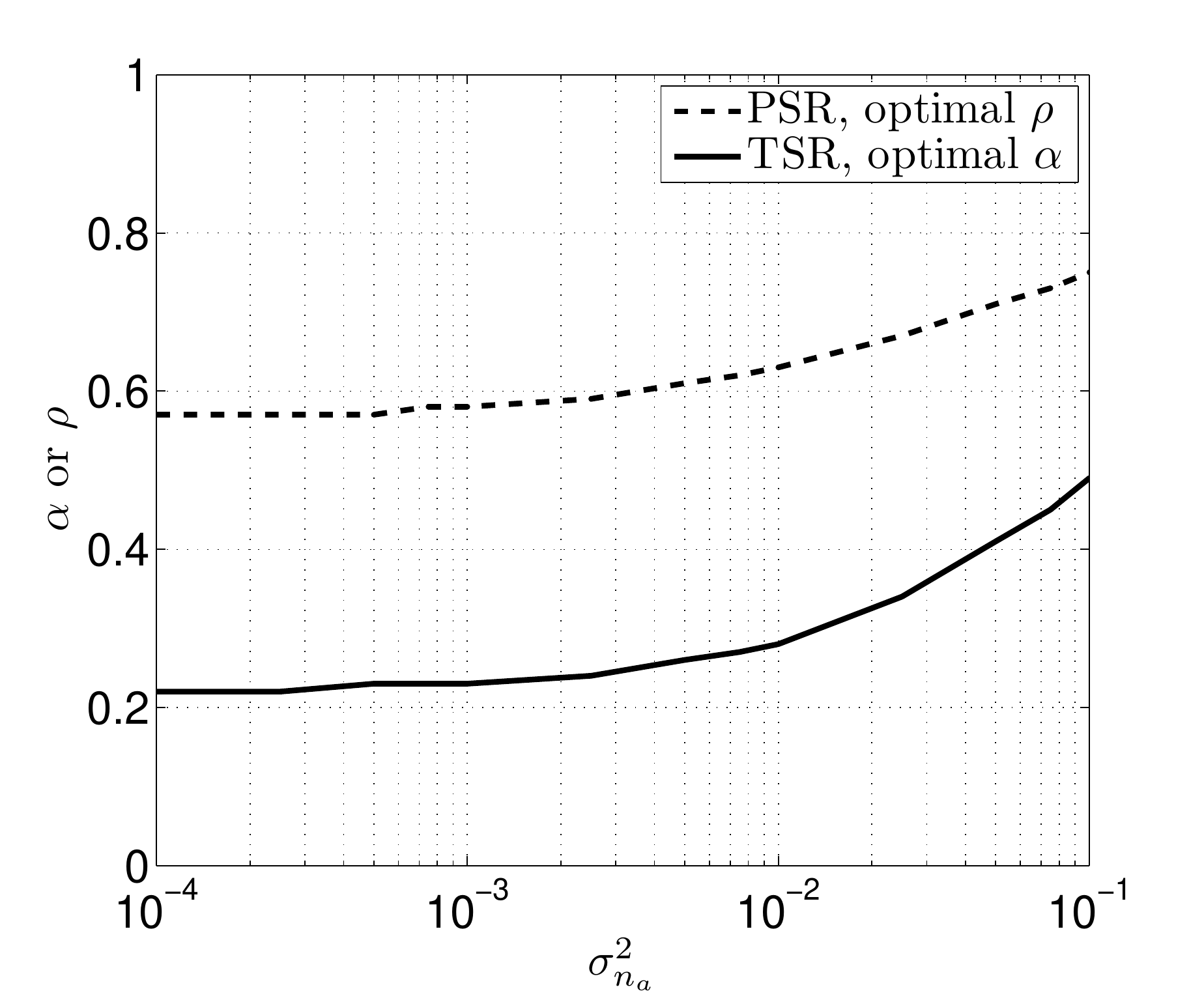}
    \label{fig:FF_noise_a}
  }
  \subfigure[]
  {
    \hspace{-0.0in}
    \includegraphics[width=0.48 \textwidth]{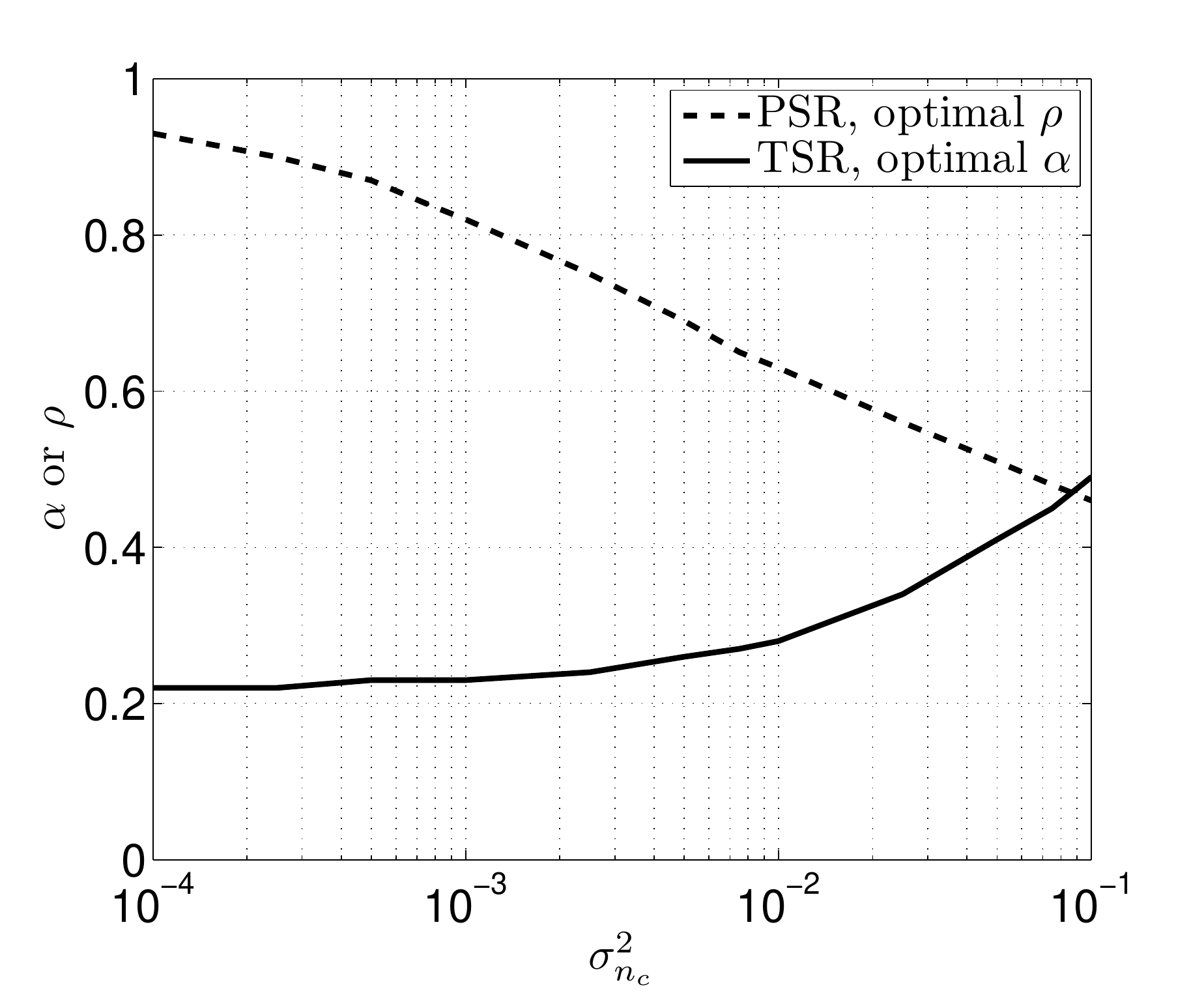}
    \label{fig:FF_noise_p}
  }
 \vspace{-0.05in}
  \caption{Optimal values of $\alpha$ and $\rho$ for the TSR and PSR protocols, respectively, in delay-limited transmission mode for (a) different values of antenna noise variance $\sigma_{n_a}^2$ and $\sigma_{n_c}^2 = 0.01$ (fixed) and (b) different values of conversion noise variance $\sigma_{n_c}^2$ and $\sigma_{n_a}^2 = 0.01$ (fixed). Other parameters: $P_s = 1$, $\eta = 1$, and $d_1 = d_2 = 1$.}
  \vspace{-0.10in}
\label{fig:FF_noise}
\end{figure*}

Fig. \ref{fig:FF_noise} plots the optimal values of $\alpha$ and $\rho$ for the TSR and the PSR protocols, respectively, in the delay-limited transmission mode for different values of antenna noise variance, $\sigma_{n_a}^2$ (see Fig. \ref{fig:FF_noise_a} for fixed $\sigma_{n_c}^2 = 0.01$) and different values of conversion noise variance, $\sigma_{n_c}^2$ (see Fig. \ref{fig:FF_noise_p} for fixed $\sigma_{n_a}^2 = 0.01$). Fig. \ref{fig:FF_noise} shows that the optimal value of $\alpha$ increases by increasing $\sigma_{n_a}^2$ or $\sigma_{n_c}^2$ However, the optimal $\rho$ increases by increasing $\siga$ (see Fig. \ref{fig:FF_noise_a}) and decreases by increasing $\sigp$ (see Fig. \ref{fig:FF_noise_p}). This is due to the fact that for the TSR protocol, both noise processes, the antenna noise at the baseband $n_a^{[r]}(k)$ and the conversion noise $n_c^{[r]}(k)$, affect the received signal $y_r(k)$ in the same way. Consequently, the trend for the optimal value of $\alpha$ is same when plotted with respect to the noise variances, $\siga$ or $\sigp$, in Fig. \ref{fig:FF_noise_a} and Fig. \ref{fig:FF_noise_p}, respectively. On the other hand, for the PSR protocol, the baseband antenna noise $n_a^{[r]}(k)$ affects the received signal $y_r(k)$ and the conversion noise $n_c^{[r]}(k)$ affects the portion of the received signal strength, $\sqrt{1-\rho} y_r(t)$ (see Fig. \ref{fig:PS_R}). As a result, the trend for the optimal value of $\rho$ is different when plotted with respect to the noise variances, $\siga$ or $\sigp$, in Fig. \ref{fig:FF_noise_a} and Fig. \ref{fig:FF_noise_p}, respectively. Similar trends for the optimal value of $\alpha$ and $\rho$ are observed in the delay-tolerant transmission mode, which are not plotted to avoid repetition.

\subsection{Effect of Relay Location}

\begin{figure}[t]
    \centering
    \includegraphics[width=0.46 \textwidth]{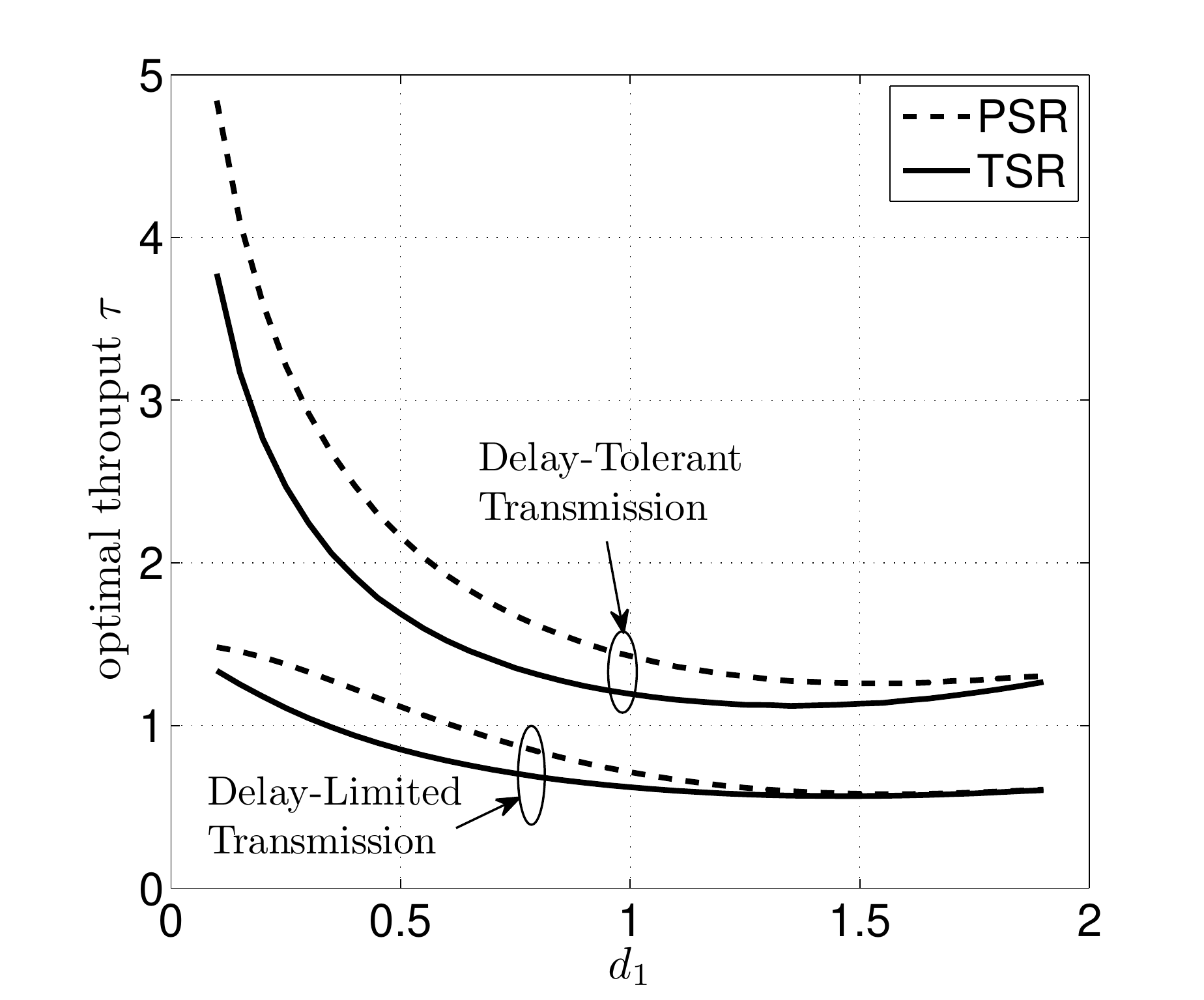} \vspace{-.15 in}
  \caption{Optimal throughput $\tau$ for the TSR and PSR protocols for different values of source to relay distance, $d_1$. Other parameters: $\sigma_{n_a}^2 = 0.01$, $\sigma_{n_c}^2 = 0.01$, $P_s = 1$, $\eta = 1$, and $d_2 = 2 - d_1$.}
    \label{fig:dist}
\end{figure}

Fig. \ref{fig:dist} plots the optimal throughput $\tau$ for the TSR and the PSR protocols in both the delay-limited and the delay-tolerant transmission modes for different values of the source to relay distance, $d_1$. The relay to destination distance, $d_2$ is set to $d_2 = 2 - d_1$ and the noise variances are kept fixed, i.e., $\sigma_{n_a}^2 = 0.01$ and $\sigma_{n_c}^2 = 0.01$. It can be observed from Fig. \ref{fig:dist} that for both the TSR and the PSR protocols, the optimal throughput $\tau$ decreases as $d_1$ increases, i.e., as the distance between source node and the relay node increases. This is because by increasing $d_1$, both energy harvested ($E_h$ defined in \eqref{eq:EH_TS} for the TSR protocol and defined in \eqref{eq:EH_PS} for the PSR protocol) and the received signal strength at the relay node ($y_r(k)$ defined in \eqref{eq:rrk_AF} for the TSR protocol and defined in \eqref{eq:rrk_AF_PS} for the PSR protocol) decrease due to the larger path loss, $d_1^m$. Consequently, the received signal strength at the destination node ($\gamma_D$ defined in \eqref{eq:gD} for the TSR protocol and defined in \eqref{eq:gD_PS} for the PSR protocol) is poor and the achievable throughput decreases. However, the throughput does not change much by increasing $d_1$ beyond $1.2$. This is because as the relay node gets closer to the destination ($d_2 < 0.8$), even lesser values of harvested energy, $E_h$ suffice for reliable communication between the relay and the destination nodes due to smaller values of the relay to destination path loss, $d_2^m$. \emph{It is important to note that, as illustrated in Fig. \ref{fig:dist}, the optimal relay location with energy harvesting is close to the source node. This is different from the general case where energy harvesting is not considered at the relay and the maximum throughput is achieved when relay is located mid-way between the source and the destination nodes.}\footnote{This is observed through simulations for non energy harvesting setup. The results, however, are not included here due to brevity.}

\begin{figure}[t]
    \centering
    \includegraphics[width=0.46 \textwidth]{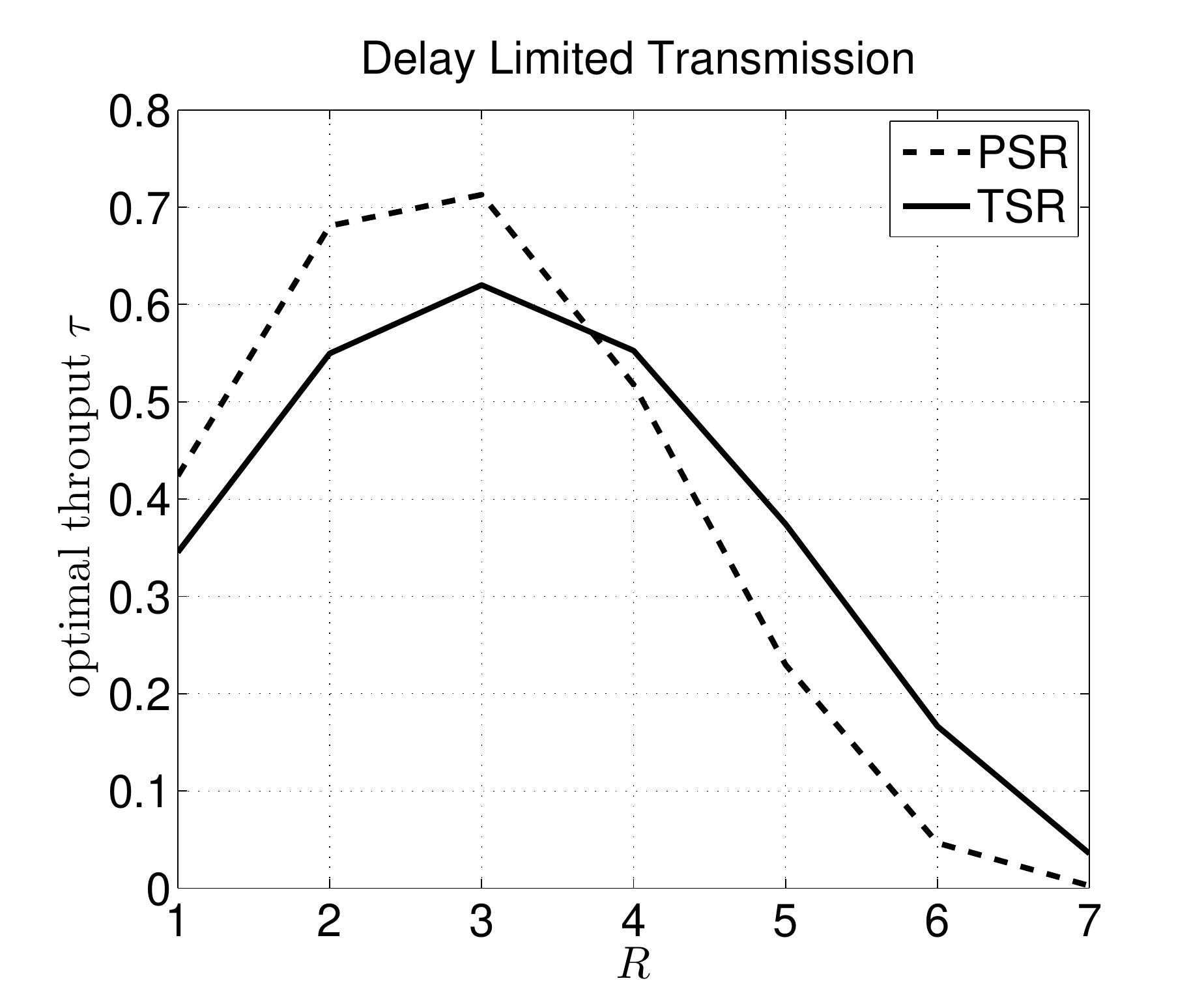} \vspace{-.15 in}
  \caption{Optimal throughput $\tau$ for the TSR and PSR protocols in delay-limited transmission mode for different values of source transmission rate, $R$. Other parameters: $\sigma_{n_a}^2 = 0.01$, $\sigma_{n_c}^2 = 0.01$, $P_s = 1$, $\eta = 1$, $d_1 = d_2 = 1$, and $\pout$ is given in \eqref{eq:pout_A1}.}
    \label{fig:R}
\end{figure}

\subsection{Effect of the Source Transmission Rate in Delay-Limited Transmission}

Fig. \ref{fig:R} plots the optimal throughput $\tau$ for the TSR and the PSR protocols in delay-limited transmission mode for different values of the source transmission rate, $R$ bits/sec/Hz. Noise variances are kept fixed, i.e., $\sigma_{n_a}^2 = 0.01$ and $\sigma_{n_c}^2 = 0.01$. Fig. \ref{fig:R} shows that the optimal $\tau$ increases as $R$ increases to a certain value but then starts decreasing for larger values of $R$. This is because the throughput depends on $R$ (see \eqref{eq:tau} for the TSR protocol and \eqref{eq:tau_PS} for the PSR protocol) and thus at relatively low transmission rates, the throughput decreases. On the other hand, for larger transmission rates $R$, the receiver fails to correctly decode the large amount of data in the limited time. Thus, the probability of outage $\pout$ increases and the throughput decreases. \emph{It can be observed from Fig. \ref{fig:R} that the PSR protocol results in more throughput than the TSR protocol at relatively low transmission rates. On the other hand, when transmitting at larger rates, the TSR protocol renders larger values of throughput compared to the PSR protocol.} Note that for the delay-tolerant transmission mode, there is no result plotted for varying the source transmission rate because the transmission rate is equal to the ergodic capacity $C$.

\begin{figure}[t]
    \centering
    \includegraphics[width=0.46 \textwidth]{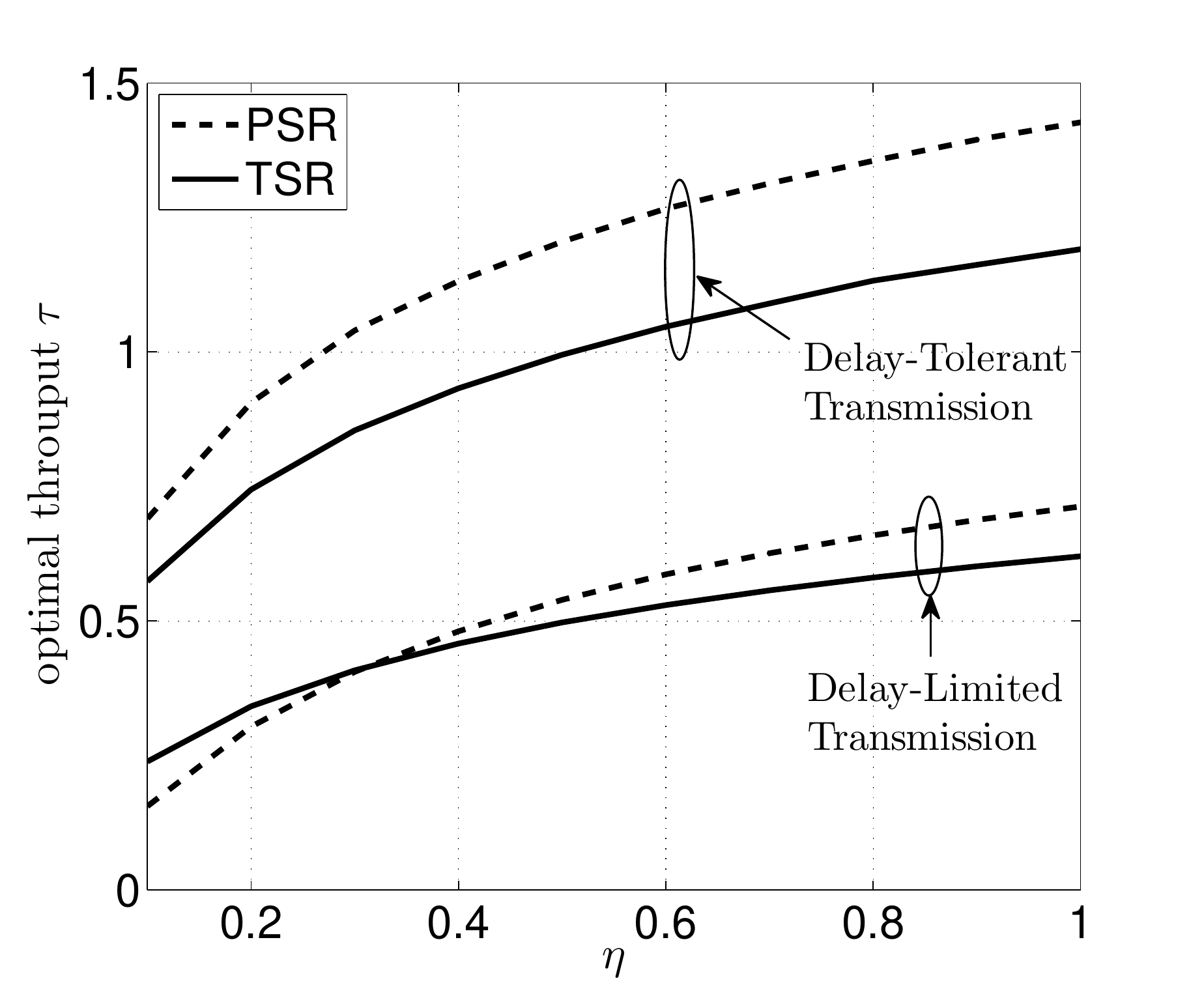} \vspace{-.15 in}
  \caption{Optimal throughput $\tau$ for the TSR and PSR protocols, respectively for different values of energy harvesting efficiency, $\eta$. Other parameters: $\sigma_{n_a}^2 = 0.01$, $\sigma_{n_c}^2 = 0.01$, $P_s = 1$, and $d_1 = d_2 = 1$.}
    \label{fig:eta}
\end{figure}

\ifCLASSOPTIONpeerreview
\else

\begin{figure*}[!b]
 \hrulefill
 \vspace{-0.0cm}
\normalsize
\setcounter{MYtempeqncnt}{26}
\begin{align}\label{eq:pout_app1}
      p_\text{out} &= p \left( \frac{2 \eta P_s^2 |h|^4 |g|^2 \alpha}{2 \eta P_s |h|^2 |g|^2 d_1^m \sigr \alpha + P_s |h|^2 d_1^m d_2^m \sigd (1- \alpha) + d_1^{2m} d_2^{m} \sigr \sigd (1-\alpha) } < \gamma_0 \right) \notag \\ &= p \left( |g|^2 <  \frac{P_s d_1^m d_2^m \sigd \gamma_0 (1 - \alpha) |h|^2 + d_1^{2m} d_2^m \sigr \sigd \gamma_0 (1 - \alpha)}{2 \eta P_s^2 \alpha |h|^4 - 2 \eta P_s d_1^m \sigr \gamma_0 \alpha  |h|^2} \right) \notag \\ &= p \left( |g|^2 < \frac{a|h|^2+b}{c |h|^4 - d|h|^2} \right) \tag{A.1}
\end{align}
\setcounter{equation}{\value{MYtempeqncnt}}
\vspace*{-.4cm}
\end{figure*}

\fi

\subsection{Effect of the Energy Harvesting Efficiency}

Fig. \ref{fig:eta} plots the optimal throughput $\tau$ for the TSR and the PSR protocols in both the delay-limited and the delay-tolerant transmission modes for different values of energy harvesting efficiency, $\eta$. \emph{Considering the delay-limited transmission mode, it can be observed from Fig. \ref{fig:eta} that for smaller values of energy harvesting efficiency $\eta$, the TSR protocol outperforms the PSR protocol in terms of throughput. On the other hand, Fig. \ref{fig:eta} shows that in the delay-tolerant transmission mode, the PSR protocol outperforms the TSR protocol for all the values of $\eta$.} Finally, Table \ref{tab:SIM} summarizes the important insights concerning the throughput comparison between the TSR and the PSR protocols in the delay-limited transmission mode. For the delay-tolerant transmission mode, the PSR protocol outperforms the TSR protocol in terms of throughput for the considered values of the various system parameters.

\ifCLASSOPTIONpeerreview

\begin{table}[t]
\caption{Throughput ($\tau$) comparison for the TSR and PSR protocols in the delay-limited transmission mode.} \centering
\begin{tabular}{|l|l|} \hline
\multicolumn{1}{|c|} {System Parameters} & \multicolumn{1}{c|} {Throughput ($\tau$) of TSR vs PSR} \\ \hline
Noise variance ($\siga$ or $\sigp$) & At low noise variance, PSR outperforms TSR and vice versa at high noise variance. \\
Source to relay distance ($d_1$) & For small values of $d_1$, PSR outperforms TSR and performance is similar for large $d_1$. \\
Transmission rate ($R$) & For small values of $R$, PSR outperforms TSR and vice versa at large values of $R$.  \\
Energy harvesting efficiency ($\eta$) & For small values of $\eta$, TSR outperforms PSR and vice versa at large values of $\eta$. \\ \hline
 \end{tabular}
\label{tab:SIM}
\end{table}

\else

\begin{table*}[t]
\vspace{1cm}
\caption{Throughput ($\tau$) comparison for the TSR and PSR protocols in the delay-limited transmission mode.} \centering
\begin{tabular}{|l|l|} \hline
\multicolumn{1}{|c|} {System Parameters} & \multicolumn{1}{c|} {Throughput ($\tau$) of TSR vs PSR} \\ \hline
Noise variance ($\siga$ or $\sigp$) & At low noise variance, PSR outperforms TSR and vice versa at high noise variance. \\
Source to relay distance ($d_1$) & For small values of $d_1$, PSR outperforms TSR and performance is similar for large $d_1$. \\
Transmission rate ($R$) & For small values of $R$, PSR outperforms TSR and vice versa at large values of $R$.  \\
Energy harvesting efficiency ($\eta$) & For small values of $\eta$, TSR outperforms PSR and vice versa at large values of $\eta$. \\ \hline
 \end{tabular}
\label{tab:SIM}
\end{table*}

\fi

\section{Conclusions}\label{sec:conclusions}

In this paper, an amplify-and-forward wireless cooperative or sensor network has been considered, where an energy constrained relay node harvests energy from the received RF signal and uses that harvested energy to forward the source signal to the destination node. Two relaying protocols, namely, i) TSR protocol and ii) PSR protocol, are proposed to enable wireless energy harvesting and information processing at the relay, based on the recently developed and widely adopted TS and PS receiver architectures. Both the delay-limited and the delay-tolerant transmission modes are considered for communication. In order to determine the achievable throughput at the destination, analytical expressions for the outage probability and the ergodic capacity are derived for the delay-limited and the delay-tolerant transmission modes, respectively. The optimal value of energy harvesting time in the TSR protocol and the optimal value of power splitting ratio in the PSR protocol are numerically investigated. The numerical analysis in this paper has provided practical insights into the effect of various system parameters on the performance of wireless energy harvesting and information processing using AF relay nodes. The key insights are summarized in Table \ref{tab:SIM}. The numerical analysis in this paper is underpinned by the derived analytical expressions for the throughput for both TSR and PSR protocols, which are summarized in Table \ref{tab:AF},

The throughput results derived in this paper represent the upper bound on the practically achievable throughput. The information theoretic work in this paper can be extended to analyze the impact of various system constraints such as finite alphabet modulation, a minimum power level required for energy harvesting, and automatic repeat requests or retransmissions in the case of packet loss, on the throughput performance. Moreover, in this work, we assume that the CSI is available at the destination node only. If the CSI is available at the relay, the proposed protocols can be made adaptive by adapting the energy harvesting time or power splitting ratio according to the channel conditions.

\appendices
\numberwithin{equation}{section}

\section{\vspace{-0pt}Proof of Proposition 1 in \eqref{eq:pout_A}}\label{app:A}
This appendix derives the $\pout$, in \eqref{eq:pout_A}, at the destination node for the TSR protocol. Substituting \eqref{eq:gD} into \eqref{eq:pout}, $\pout$ is given by
\ifCLASSOPTIONpeerreview
\begin{align}\label{eq:pout_app1}
      p_\text{out} &= p \left( \frac{2 \eta P_s^2 |h|^4 |g|^2 \alpha}{2 \eta P_s |h|^2 |g|^2 d_1^m \sigr \alpha + P_s |h|^2 d_1^m d_2^m \sigd (1- \alpha) + d_1^{2m} d_2^{m} \sigr \sigd (1-\alpha) } < \gamma_0 \right) \notag \\ &= p \left( |g|^2 <  \frac{P_s d_1^m d_2^m \sigd \gamma_0 (1 - \alpha) |h|^2 + d_1^{2m} d_2^m \sigr \sigd \gamma_0 (1 - \alpha)}{2 \eta P_s^2 \alpha |h|^4 - 2 \eta P_s d_1^m \sigr \gamma_0 \alpha  |h|^2} \right) \notag \\ &= p \left( |g|^2 < \frac{a|h|^2+b}{c |h|^4 - d|h|^2} \right)
\end{align}
\else
\addtocounter{equation}{1}
\eqref{eq:pout_app1} at the bottom of the page,
\fi
where $a \triangleq P_s d_1^m d_2^m \sigd \gamma_0 (1 - \alpha) $, $b \triangleq d_1^{2m} d_2^m \sigr \sigd \gamma_0 (1 - \alpha) $, $c \triangleq 2 \eta P_s^2 \alpha $, and $d \triangleq 2 \eta P_s d_1^m \sigr \gamma_0 \alpha $. Given the factor in the denominator, $c |h|^4 - d|h|^2$, can be positive or negative, $\pout$ is given by
\begin{align}\label{eq:pout_app2}
      p_\text{out} &= p \left( (c |h|^4 - d|h|^2) |g|^2 < (a|h|^2+b) \right) \notag \\
       &= \begin{cases}
                    p \left( |g|^2 < \frac{a|h|^2+b}{c |h|^4 - d|h|^2} \right) , & |h|^2 < d/c \\
                    p \left( |g|^2 > \frac{a|h|^2+b}{c |h|^4 - d|h|^2} \right) = 1 , & |h|^2 > d/c
       \end{cases}
\end{align}
The second equality in \eqref{eq:pout_app2} follows due to the fact that if $|h|^2 > d/c$, $c |h|^4 - d|h|^2$ will be a negative number and probability of $|g|^2$ being greater than some negative number is always $1$. Following \eqref{eq:pout_app2}, $\pout$ is given by
\ifCLASSOPTIONpeerreview
\begin{align}\label{eq:pout_app22}
p_\text{out} &= \int_{z=0}^{d/c} f_{|h|^2}(z) p \left( |g|^2 > \frac{az+b}{c z^2 - dz}  \right)  dz + \int_{z=d/c}^{\infty} f_{|h|^2}(z) p \left( |g|^2 < \frac{az+b}{c z^2 - dz}  \right)  dz \notag \\
&=  \int_{z=0}^{d/c} f_{|h|^2}(z)   dz + \int_{z=d/c}^{\infty} f_{|h|^2}(z) \left( 1 - e^{- \frac{az+b}{(cz^2-dz) \lgg }} \right)  dz
\end{align}
\else
\begin{align}\label{eq:pout_app22}
p_\text{out} &= \int_{z=0}^{d/c} f_{|h|^2}(z) p \left( |g|^2 > \frac{az+b}{c z^2 - dz}  \right)  dz  \notag \\ & \hspace{2.3cm} + \int_{z=d/c}^{\infty} f_{|h|^2}(z) p \left( |g|^2 < \frac{az+b}{c z^2 - dz}  \right)  dz \notag \\
&=  \int_{z=0}^{d/c} f_{|h|^2}(z)   dz + \int_{z=d/c}^{\infty} f_{|h|^2}(z) \left( 1 - e^{- \frac{az+b}{(cz^2-dz) \lgg }} \right)  dz
\end{align}
\fi
where $z$ is the integration variable, $f_{|h|^2}(z) \triangleq \frac{1}{\lh} e^{-z/\lh}$ is the probability density function (PDF) of exponential random variable $|h|^2$, $\lh$ is the mean of the exponential random variable $|h|^2$, $ F_{|g|^2}(z) \triangleq p ( |g|^2 < z ) = 1 - e^{-z/\lgg}$ is the cumulative distribution function (CDF) of the exponential random variable $|g|^2$ and $\lgg$ is the mean of the exponential random variable $|g|^2$. Substituting $f_{|h|^2}(z) = \frac{1}{\lh} e^{-z/\lh}$ in \eqref{eq:pout_app22}, $\pout$ is given by
\begin{align}\label{eq:pout_app3}
p_\text{out} = 1 - \frac{1}{\lambda_h} \int_{z = d/c}^{\infty}  e^ {- \left( \frac{z}{\lambda_h} + \frac{az+b}{(cz^2-dz)\lambda_g}   \right)  } dz
\end{align}
\eqref{eq:pout_app3} presents the analytical expression of $p_\text{out}$ for the TSR protcol, as presented in Proposition 1 in \eqref{eq:pout_A}.

The integration in \eqref{eq:pout_app3} cannot be further simplified. However, one can apply a high SNR approximation and obtain further simplified expression for $p_\text{out}$. At high SNR, the third factor in the denominator of \eqref{eq:gD}, $ d_1^{2m} d_2^{m} \sigr \sigd (1-\alpha)  $, is negligible (because of the product of the two noise variance terms) compared to the other two factors in the denominator, $2 \eta P_s |h|^2 |g|^2 d_1^m \sigr \alpha$ and $P_s |h|^2 d_1^m d_2^m \sigd (1- \alpha) $, i.e., $\gamma_D \approx \frac{2 \eta P_s^2 |h|^4 |g|^2 \alpha}{2 \eta P_s |h|^2 |g|^2 d_1^m \sigr \alpha + P_s |h|^2 d_1^m d_2^m \sigd (1- \alpha)  }$. In other words, at high SNR, the constant $b$ can be approximated by $0$, i.e., $b = d_1^{2m} d_2^m \sigr \sigd \gamma_0 (1 - \alpha) \approx 0 $. Thus, $p_\text{out} $ in \eqref{eq:pout_app3} can be approximated as
\begin{align}\label{eq:pout_app4}
p_\text{out} \approx 1 - \frac{1}{\lambda_h} \int_{z = d/c}^{\infty}  e^ {- \left( \frac{z}{\lambda_h} + \frac{a}{(cz-d)\lambda_g}   \right)  } dz
\end{align}
Let us define a new integration variable $x \triangleq cz-d$. Thus, approximated outage at high SNR is given by %
\begin{align}\label{eq:pout_app5}
p_\text{out} &\approx 1 -  \frac{e^{- \frac{d}{c \lambda_h}}}{c \lambda_h} \int_{x=0}^{\infty}  e^ {- \left( \frac{x}{\lambda_h c} + \frac{a}{x \lgg}   \right)  } dx \notag \\ &= 1 - e^{- \frac{d}{c \lambda_h}} u K_{1} \left(  u \right)
\end{align}
where $u \triangleq \sqrt{\frac{4a}{c \lh \lgg} }$, $K_{1}(\cdot)$ is the first-order modified Bessel function of the second kind \cite{Gradshteyn-80-B} and the last equality is obtained by using the formula, $\int_{0}^{\infty} e^{-\frac{\beta}{4x} - \gamma x} dx = \sqrt{\frac{\beta}{\gamma}} K_{1}( \sqrt{\beta \gamma } )$ \cite[$\S$3.324.1]{Gradshteyn-80-B}. This ends the proof for Proposition 1.

\section{\vspace{-0pt}Proof of Proposition 2 in \eqref{eq:C_A}}\label{app:B}

In order to find the analytical expression for the ergodic capacity, the PDF of $\gamma_D$, $f_{\gamma_D}(\gamma)$, needs to be evaluated first. The PDF of $\gamma_D$ can be obtained from the CDF of $\gamma_D$, $F_{\gamma_D}(\gamma)$ which is given by
\begin{align}\label{eq:Fgamma}
      F_{\gamma_D}(\gamma) =  p(  \gamma_D < \gamma )
       = 1 - \displaystyle\frac{1}{\lambda_h} \int_{z = d/c}^{\infty}  e^ {- \left( \frac{z}{\lambda_h} + \frac{az+b}{(cz^2-dz)\lambda_g}   \right)  } dz,
\end{align}
where $a \triangleq P_s d_1^m d_2^m \sigd \gamma (1 - \alpha) $, $b \triangleq d_1^{2m} d_2^m \sigr \sigd \gamma (1 - \alpha) $, $c \triangleq 2 \eta P_s^2 \alpha $, $d \triangleq 2 \eta P_s d_1^m \sigr \gamma \alpha $, and equality in \eqref{eq:Fgamma} follows from \eqref{eq:pout} and \eqref{eq:pout_A1}. Using \eqref{eq:Fgamma}, the PDF of $\gamma_D$ is given by
\ifCLASSOPTIONpeerreview
\begin{align}\label{eq:fgamma}
      f_{\gamma_D}(\gamma) =  \frac{\partial F_{\gamma_D}(\gamma)}{\partial \gamma}
       = \frac{1}{\lambda_h \gamma} \int_{z = d/c}^{\infty} \frac{(az+b) c z^2}{(cz^2-dz)^2 \lambda_g}  e^ {- \left( \frac{z}{\lambda_h} + \frac{az+b}{(cz^2-dz)\lambda_g}   \right)  } dz.
\end{align}
\else
\begin{align}\label{eq:fgamma}
      f_{\gamma_D}(\gamma) &=  \frac{\partial F_{\gamma_D}(\gamma)}{\partial \gamma} \notag \\
       &= \frac{1}{\lambda_h \gamma} \int_{z = d/c}^{\infty} \frac{(az+b) c z^2}{(cz^2-dz)^2 \lambda_g}  e^ {- \left( \frac{z}{\lambda_h} + \frac{az+b}{(cz^2-dz)\lambda_g}   \right)  } dz.
\end{align}
\fi
Using \eqref{eq:C1} and the PDF $f_{\gamma_D}(\gamma)$ in \eqref{eq:fgamma}, the ergodic capacity $C$ is given by
\begin{subequations}\label{eq:C_app1}
\ifCLASSOPTIONpeerreview
\begin{align}
      C &= \int_{\gamma = 0}^{\infty} f_{\gamma_D}(\gamma) \log_2(1+\gamma) d\gamma  \label{eq:C_app1a} \\
       &= \int_{\gamma = 0}^{\infty} \int_{z = d/c}^{\infty}  \frac{(az+b) c z^2}{(cz^2-dz)^2 \lambda_g \lambda_h \gamma} e^ {- \left( \frac{z}{\lambda_h} + \frac{az+b}{(cz^2-dz)\lambda_g}   \right)  }  \log_2(1 + \gamma)  dz d \gamma. \label{eq:C_app1b}
\end{align}
\else
\begin{align}
      C &= \int_{\gamma = 0}^{\infty} f_{\gamma_D}(\gamma) \log_2(1+\gamma) d\gamma  \label{eq:C_app1a} \\
       &= \int_{\gamma = 0}^{\infty} \int_{z = d/c}^{\infty}  \frac{(az+b) c z^2}{(cz^2-dz)^2 \lambda_g \lambda_h \gamma} e^ {- \left( \frac{z}{\lambda_h} + \frac{az+b}{(cz^2-dz)\lambda_g}   \right)  } \notag \\ & \hspace{4.5cm}  \log_2(1 + \gamma)  dz d \gamma. \label{eq:C_app1b}
\end{align}
\fi
\end{subequations}
\eqref{eq:C_app1b} presents the analytical expression of $C$ for the TSR protocol, as presented in Proposition 2 in \eqref{eq:C_A}.

The integration in \eqref{eq:C_app1b} cannot be further simplified. However, one can apply high SNR approximation, as explained earlier in Appendix \ref{app:A} below \eqref{eq:pout_app3}, to further simplify the expression in \eqref{eq:C_app1b}. Thus, using \eqref{eq:Fgamma}, the approximate value for the CDF of $\gamma_D$ is given by
\begin{align}\label{eq:Fgamma_app}
      F_{\gamma_D}(\gamma) &\approx  1 - \frac{1}{\lambda_h} \int_{z = d/c}^{\infty}  e^ {- \left( \frac{z}{\lambda_h} + \frac{a}{(cz-d)\lambda_g}   \right)  } dz \notag \\ &= 1 - e^{- \frac{d}{c \lambda_h}} u K_1 \left(  u \right),
\end{align}
where $u \triangleq \sqrt{\frac{4a}{c \lh \lgg} }$,$a$, $c$, and $d$ are defined below \eqref{eq:Fgamma} and the second equality in \eqref{eq:Fgamma_app} follows using \eqref{eq:pout_app4}-\eqref{eq:pout_app5}. Evaluating the derivative of $F_{\gamma_D}(\gamma)$ in \eqref{eq:Fgamma_app} with respect to $\gamma$, the PDF of $\gamma_D$ can be approximated as \vspace{-0pt}
\begin{align}\label{eq:fgamma_app}
      f_{\gamma_D}(\gamma) &\approx  \frac{u^2 K_0(u) e^{- \frac{d}{c \lambda_h}} }{2 \gamma} + \frac{ d u K_1(u) e^{- \frac{d}{c \lambda_h}} }{ \gamma c \lambda_h}
\end{align}
where \eqref{eq:fgamma_app} follows from \eqref{eq:Fgamma_app} using the property of Bessel function, $ \frac{d}{dz} \left( z^v K_v(z) \right) = - z^v K_{v-1}(z)$ \cite[$\S$8.486.18]{Gradshteyn-80-B}. Thus, using \eqref{eq:C_app1a} and \eqref{eq:fgamma_app}, approximated ergodic capacity at high SNR is given by
\begin{align}\label{eq:C_app_approx}
C  \approx  \displaystyle\int_{\gamma = 0}^{\infty} \left( \frac{u^2 K_0(u) e^{- \frac{d}{c \lambda_h}} }{2 \gamma} + \frac{ d u K_1(u) e^{- \frac{d}{c \lambda_h}} }{ \gamma c \lambda_h}  \right) \log_2(1 + \gamma)  d \gamma
\end{align}
This ends the proof for Proposition 2.

\ifCLASSOPTIONpeerreview
\vspace{-0.5cm}
\else
\fi


\end{document}